%% file: main.tex
\documentclass[sigplan,screen,nonacm]{acmart}

\pdfoutput=1

\copyrightyear{2022}
\acmYear{2022}
\setcopyright{acmlicensed}
\acmConference[LICS '22]{37th Annual ACM/IEEE Symposium on Logic in Computer Science (LICS)}{August 2--5, 2022}{Haifa, Israel}
\acmBooktitle{37th Annual ACM/IEEE Symposium on Logic in Computer Science (LICS) (LICS '22), August 2--5, 2022, Haifa, Israel}
\acmPrice{15.00}
\acmDOI{10.1145/3531130.3533370}
\acmISBN{978-1-4503-9351-5/22/08}

\input{preamble}

\title{Concrete categories and higher-order recursion}
\subtitle{With applications including probability, differentiability, and full abstraction}

\author{Cristina Matache}
\affiliation{%
  \institution{University of Oxford}
  \department{Department of Computer Science}
  \city{Oxford}
  \country{UK}
}
\author{Sean Moss}
\affiliation{%
  \institution{University of Oxford}
  \department{Department of Computer Science}
  \city{Oxford}
  \country{UK}
}
\author{Sam Staton}
\affiliation{%
  \institution{University of Oxford}
  \department{Department of Computer Science}
  \city{Oxford}
  \country{UK}
}

\begin{document}

\begin{abstract}
  We study concrete sheaf models for a call-by-value higher-order language with recursion. Our family of sheaf models is a generalization of many examples from the literature, such as models for probabilistic and differentiable programming, and fully abstract logical relations models. We treat recursion in the spirit of synthetic domain theory. We provide a general construction of a lifting monad starting from a class of admissible monomorphisms in the site of the sheaf category. In this way, we obtain a family of models parametrized by a concrete site and a class of monomorphisms, for which we prove a general computational adequacy theorem.
\end{abstract}



\maketitle

\section{Introduction}
\input{intro}

\section*{\textsc{Part I: Concrete sheaf models of PCF}}

\input{concsheaves}

\input{cpo-with-functions}

\input{language}
\input{adequacy}

\section*{\textsc{Part II: Understanding models through synthetic domain theory}}
In this second part of the paper, we explain why our requirements on the class of admissible monos (Def.~\ref{def:class-of-admissible-monos}) are canonical, by demonstrating how they arise generally from synthetic domain theory. We treat partiality (\S\ref{sec:beginning-of-sdt})  and recursion (\S\ref{sec:recursion}) separately, before explaining $\omega$-concrete sheaves from this perspective (\S\ref{sec:sheaf-model-pcfv}). 
\section{Partiality in categories of sheaves via dominances and pre-admissible monos}
\input{dominance}
\input{sheaves}

\input{recursion}

\input{sheaf-model}

\subsection{Connection to synthetic domain theory}
The general adequacy theorem (Thm.~\ref{thm:sheaves-adequate}) connects to the synthetic/axiomatic domain theory literature on general adequacy theorems, for example by Fiore and Plotkin~\cite{fiore-plotkin-adequacy,fiore_1996} and Simpson~\cite{simpson-computational-adequacy-in-an-elementary-topos,DBLP:journals/apal/Simpson04}.
In particular, our model in the sheaf topos $\sheaves(\bC+\vertendo_0,J\cup J_\vseq)$ is an instance of Simpson's more general \emph{natural model of synthetic domain theory}~\cite{simpson-computational-adequacy-in-an-elementary-topos}, that is, an elementary topos with a dominance and a natural numbers object which is well-complete, and
our adequacy result (Thm.~\ref{thm:sheaves-adequate}) follows from~\cite[Thm.~2]{simpson-computational-adequacy-in-an-elementary-topos}.
To see this, notice that any non-trivial Grothendieck topos is $1$-consistent;
the dominance there is used to construct a lifting monad in the same way as we do in~\Cref{sec:dominance};
the initial algebra $\mathbf{I}$ and final coalgebra~$\mathbf{F}$ there play the role of $\omega$ and $\wbar$; completeness is defined similarly and is used to prove a fixed point theorem \cite[Prop.~2]{simpson-computational-adequacy-in-an-elementary-topos} corresponding to our~\Cref{thm:fixed-points-x}.
Thus from this perspective, our contribution here is a method for obtaining a topos with a dominance (via the concrete site and the class of admissible monos, Def.~\ref{def:class-of-admissible-monos}) such that the natural numbers object is necessarily well-complete (Prop.~\ref{prop:lnat-complete}).

We also note that Sterling and Harper have developed another interesting perspective on adequacy for sheaf-based models, e.g.~in~\cite{sterling-harper-sheaf-semantics-of-termination-insensitive-noninterference}, and we expect that our adequacy proof could be rephrased to fit into their framework.

\section{Summary}
In Part~I, we presented an elementary framework for building semantic models of functional programming languages. The key ingredients are a concrete site (Def.~\ref{def:omnibus-concrete}) and a class of admissible monos (Def.~\ref{def:class-of-admissible-monos}). In Part~II, we explained how our requirements on the class of admissible monos are canonical in that they connect to general constructions from synthetic domain theory. Such a semantic model is necessarily adequate (Thms~\ref{thm:wconc-adequate},\ref{thm:sheaves-adequate}). The framework covers numerous examples from the literature (\S~\ref{sec:examplesconc},\ref{sec:exampleswconc}).

\paragraph{Acknowledgements} We have benefited from discussing this work with numerous people, including Ohad Kammar, Alex Lew, Matthijs V\'ak\'ar, Hongseok Yang, the Oxford group, and anonymous reviewers. The material is based upon work supported by an EPSRC studentship; Balliol College, Oxford;
Clarendon Fund scholarships; a Junior Research Fellowship at University College, Oxford; AFOSR award number FA9550-21-1-0038; the ERC BLAST grant; and a Royal Society University Research Fellowship.

\bibliographystyle{abbrv}
\bibliography{main}

\clearpage
\appendix
\appendixpage
\input{app-typing}

\end{document}

%% file: preamble.tex
\usepackage[utf8]{inputenc}
\usepackage[UKenglish]{babel}
\usepackage{mathtools}
\usepackage{amsthm}
\usepackage{enumitem}
\usepackage{mathpartir}
\usepackage{appendix}
\usepackage{stmaryrd}

\usepackage{stfloats}

\usepackage{tikz}
\usetikzlibrary{cd}
\usetikzlibrary{calc}

\usepackage{cleveref}
\usepackage{todonotes}
\definecolor{mycolour1}{HTML}{C9D5AE}
\definecolor{mycolour2}{HTML}{AED5CE}
\definecolor{mycolour3}{HTML}{BAAED5}
\definecolor{mycolour4}{HTML}{D5AEB6}

\DeclareMathOperator{\dom}{dom}

\newcommand{\midd}{\mathrel{\big|}}

\newcommand{\op}{\mathrm{op}}
\newcommand{\Set}{\mathsf{Set}}
\newcommand{\sets}{\Set}
\newcommand{\im}{\mathsf{im}}
\newcommand{\ssp}{\ensuremath{\mathsf{SSP}}}

\newcommand{\vertendo}{\mathbb V} 
\newcommand{\vsets}{\mathsf{vSet}}

\newcommand{\vseq}{\mathrm{V}}
\newcommand{\wbar}{{\overline{\omega}}}
\newcommand{\succe}{\mathsf{succ}} 

\newcommand{\logval}[1]{\triangleleft^{\mathsf{v}}_{#1}}
\newcommand{\logcomp}[1]{\triangleleft^{\mathsf{c}}_{#1}}

\newcommand{\wcpo}{\ensuremath{\omega\mathsf{CPO}}}

\newcommand{\om}{\mathcal O_\MCM}

\newcommand{\omegaConc}{\omega\mathsf{Conc}}

\newcommand{\pC}{\mathrm{p}\bC}
\newcommand{\pE}{\mathrm{p}\bE}

\newcommand{\bC}{\ensuremath{\mathbb C}}

\newcommand{\bE}{\ensuremath{\mathbb E}}

\newcommand{\bN}{\ensuremath{\mathbb N}}

\newcommand{\bR}{\ensuremath{\mathbb R}}

\newcommand{\MCC}{\ensuremath{\mathcal C}}

\newcommand{\MCE}{\ensuremath{\mathcal E}}

\newcommand{\MCG}{\ensuremath{\mathcal G}}
\newcommand{\MCH}{\ensuremath{\mathcal H}}
\newcommand{\MCI}{\ensuremath{\mathcal I}}

\newcommand{\MCM}{\ensuremath{\mathcal M}}

\newcommand{\sub}{\ensuremath{\mathsf{Sub}}}
\newcommand{\sheaves}{\ensuremath{\mathsf{Sh}}}
\newcommand{\presh}[1]{\ensuremath{\mathsf{PSh}(#1)}}
\newcommand{\nattrans}{\ensuremath{\mathsf{Nat}}}
\newcommand{\conc}{\ensuremath{\mathsf{Conc}}}

\DeclarePairedDelimiter{\abs}{\lvert}{\rvert}

\newcommand{\pcfv}{\ensuremath{\mathsf{PCF_v}}}

\newcommand{\cterm}{\mathrel{\vdash^{\mathbf{c}}}}
\newcommand{\vterm}{\mathrel{\vdash^{\mathbf{v}}}}
\newcommand{\nat}{\mathsf{nat}}
\newcommand{\treal}{\mathsf{real}}
\newcommand{\tzero}{\mathsf{0}}
\newcommand{\tone}{\mathsf{1}}
\newcommand{\rec}[3]{\mathsf{rec}\,#1\,#2.\,#3}
\newcommand{\letin}[3]{\mathsf{let}\, #1=#2\, \mathsf{in}\, #3}
\newcommand{\inl}[1]{\mathsf{inl}\,#1}
\newcommand{\inr}[1]{\mathsf{inr}\,#1}
\newcommand{\vzero}{\underline{\mathsf{0}}}
\newcommand{\suc}[1]{{\mathsf{S}}(#1)}
\newcommand{\lbd}[2]{\lambda #1.\,#2}
\newcommand{\ret}[1]{\mathsf{return}\,#1}
\newcommand{\casesum}[5]{\mathsf{case}\, #1\, \mathsf{of}\, \{\mathsf{inl}\, #2 \rightarrow #3,\ \mathsf{inr}\, #4\rightarrow #5\}}
\newcommand{\casenat}[4]{\mathsf{case}\, #1\ \mathsf{of}\ \{\vzero{\shortrightarrow} #2,\, \suc{#3}{\shortrightarrow} #4\} }
\newcommand{\caseempty}[1]{\mathsf{case}\, #1\, \mathsf{of}\ \{\}}

\newcommand{\den}[1]{\llbracket #1\rrbracket}

\AtEndPreamble{
  \theoremstyle{acmdefinition}
  \newtheorem{remark}[theorem]{Remark}
}


%% file: intro.tex
This paper is about semantic models of functional programming languages.
A widely accepted model involves interpreting types as chain-complete partial orders and
programs as continuous maps.
Since programs involving recursion might not terminate, it is more accurate to say that programs
are interpreted as continuous \emph{partial} maps with admissible domain.
While this is a useful interpretation, in many circumstances the literature suggests a more refined characterization of the kinds of partial map that we use to interpret programs. 
For example,
\begin{itemize}
\item In the quest for a fully abstract semantics, we would like to focus on partial maps
  that are definable \cite{DBLP:journals/iandc/OHearnR95,DBLP:journals/iandc/RieckeS02a,matache-moss-staton-fscd-2021}.
\item In probabilistic programming, we would like to interpret programs as partial maps that are Borel and with Borel domain, so that we can use Lebesgue integration to find expected values~(e.g.~\cite{DBLP:journals/pacmpl/VakarKS19}).
\item In differentiable programming and automatic differentiation, we would focus on partial maps that are smooth and with domain an open set, so that we can calculate gradients~\cite{DBLP:conf/esop/Vakar21}.
\item In a more sophisticated setting, we might insist on smoothness except for a well-behaved collection of discontinuities~\cite{DBLP:journals/corr/abs-2111-15456}.
\end{itemize}
There are further examples: in some circumstances we might require functions to be sequentially continuous on a specified domain (e.g.~\cite{DBLP:conf/esop/BartheCLG20}), in quantum programming we would require functions between spaces of density matrices to be completely positive, and so on. 

Note that these kinds of question are non-trivial. If we are only interested in, say, programming smooth functions $\treal\to\treal$, we might still use higher order functions and recursion as part of our program, and so the challenge is to show that despite these other language features, the definable
functions still amount to smooth maps. 
In this paper we give a general framework for exploring these kinds of problem, which explains this prior work on developing models for the above application domains and suggests new application domains too (\S\ref{sec:examplesconc},\ref{sec:exampleswconc}).

We emphasise that each of these application domains comes with important specific issues not covered by the general framework. For example, in probabilistic programming one would have extra features for Monte Carlo simulation, in differentiable programming one would find automatic differentiation macros, and so on. The point of our work is that we elicit a uniform foundation for building the semantic models used in all of these different applications.

\medskip 

In the remainder of this introduction, we summarize the main development of our paper at a high level. The key idea is that we use methods from synthetic domain theory to find elementary and convenient notions of partial maps (\S\ref{intro:sdt}) in concrete categories of sheaves (\S\ref{intro:concshf}), so as to obtain a general framework that provides an adequacy theorem (Thms.~\ref{thm:wconc-adequate},\ref{thm:sheaves-adequate}) for these different application domains.

\subsection{Concrete categories and sheaves}
\label{intro:concshf}
The basic setting of this paper is that we interpret each type of our programming language as a set with structure, and each typed program as a function with certain properties.
The theory of \emph{concrete categories} (Def.~\ref{def:omnibus-concrete}) is a general formalization of this situation of sets with structure and functions between them. A concrete category~$\bC$ comprises a collection of objects, with each object $c$ associated to a set $\abs c$, and then we specify which functions $\abs c\to \abs d$ are allowed as morphisms $c\to d$. For example, we have a concrete category of chain complete partial orders.

Concreteness connects with the idea of extensionality in programming language semantics.
If two programs are interpreted as different morphisms then these morphisms are actually different functions between sets and so we can distinguish them by simply applying them to different values. 

Our focus in this work is on the method of concrete sheaves, which is a method for building concrete categories that support function types and so are convenient for programming language semantics.
If the reader is familiar with logical relations, concrete sheaves can be regarded as, roughly, reflexive logical relations of varying arity. 
Categories of concrete sheaves are determined by sites~(Def.~\ref{def:omnibus-concrete}).
A first example, corresponding to a particularly simple site, is the category of sets $\abs X$ equipped with reflexive binary relations $R\subseteq \abs X \times \abs X$ (Ex.~\ref{ex:reflrel}).
For a more elaborate example, we consider diffeological spaces, a general model of smoothness (Ex.~\ref{ex:smooth}).
These are sets $\abs X$ equipped with a family of relations $R^U\subseteq [U\rightarrow\abs X]$, one for each open subset $U$ of each Euclidean space $\bR^n$. Note that the arity of these relations is typically uncountable.
For example, the tuples in  $R^\bR\subseteq [\bR\rightarrow\abs X]$ are thought of as the `smooth curves' $\bR\to \abs X$. The terminology `sheaf' refers to a gluing condition, which says for example that if we have a function $f:\bR\to \abs X$ such that the restrictions $f|_{(-\infty,1)}:(-\infty,1)
  \to \abs X$ and $f|_{(0,\infty)}:(0,\infty)
  \to \abs X$ are smooth curves in $R^{(-\infty,1)}$ and $R^{(0,\infty)}$ respectively, then $f$ itself must be regarded as smooth curve in $R^\bR$ (Def.~\ref{def:concr-sheav}). This sheaf condition constrains the colimit structure in the category, which in turn affects the interpretation of colimit types such as the natural numbers. 

  Categories of concrete sheaves are convenient for higher order languages because we can interpret the base types (such as $\treal$ and $\nat$) and then the function spaces are well behaved and straightforward to calculate (\S\ref{sec:denot-semant-pcfv}).

  \subsection{Partiality, lifting and admissible monos}
  \label{intro:sdt}
The main novelty of our paper is in our general treatment of partiality and recursion in categories of concrete sheaves. To obtain this in a canonical way, we pass to the very general framework of `synthetic domain theory', building on a long tradition (e.g.~\cite{rosolini-phd,longley-simpson-sdt-real}), and then bring this to bear on categories of concrete sheaves, extracting elementary criteria for adequate models (connecting to e.g.~\cite{fiore-plotkin-adequacy,simpson-computational-adequacy-in-an-elementary-topos}). 

\subsubsection*{Dominances and completeness}
Synthetic domain theory can be thought of as taking place within a model  of intuitionistic set theory (formally, a topos). For the approach to recursion in our development there are two key steps.
First, we should identify a \emph{dominance}, which is (informally for now) an object of semi-decidable truth values (\S\ref{sec:dominance}). This dominance induces a notion of partial function and a notion of lifting $(-)_\bot$, so that to give a partial map $X\rightharpoonup Y$ is to give a total map $X\to Y_\bot$. Lifting forms a monad, so we can interpret programs involving partiality using Moggi's method~\cite{moggi-metalanguage}. 

Second, from the dominance we build two objects which can be thought of as chains: $\omega$ and $\wbar$ (\S\ref{sec:recursion}). Intuitively, $\omega$ is an internal object describing the vertical natural numbers $\{0\leq 1\leq 2\leq \dots\}$, and $\wbar$ is the completed vertical natural numbers $\{0\leq 1\leq 2\leq \dots\infty\}$. We say that an object~$X$ is \emph{complete}, informally, if every chain $\omega\to X$ can be converted to a completed chain $\wbar\to X$ (Def.~\ref{def:complete}). We have a general treatment of recursion for complete objects, based on Tarski's fixed point theorem (Thm.~\ref{thm:fixed-points-x}). We can then give an interpretation for a programming language, provided all type constructions are interpreted as complete objects.

\subsubsection*{From concrete sheaves to synthetic domain theory}

The general framework of synthetic domain theory works well in a topos, in particular in a category of sheaves, and more generally we can restrict to just the concrete sheaves.
As a recipe for building such categories with sufficient supply of complete objects, we follow~\cite{fiore-rosolini-h,fiore-rosolini-2sdt} in considering specifically the partial order $\vseq=\{0\leq 1\leq \dots \leq \infty\}$. (This is not to be confused with $\wbar$, which is an internal construction.) 
We consider a specific category of concrete sheaves, concrete v-sets, which are sets $\abs X$ equipped with a given set $R\subseteq [\vseq\rightarrow\abs X ]$ of chains with least upper bounds (satisfying conditions, see \S\ref{sec:vset}).
For example, any chain-complete partial order determines a concrete v-set, and the relation-preserving maps are continuous functions. Tarski's fixed point theorem for chain-complete partial orders can be regarded as actually a fixed point theorem for concrete v-sets that are complete in the sense of synthetic domain theory. 
We can then straightforwardly combine this site $\vseq$ for chain-complete partial orders with any other site $\bC$  (Lem.~\ref{lem:comb-sites}), such as the site for probabilistic programming, or the site for differentiable programming, or the site for full definability.

To interpret recursion, all that remains is to find a dominance for this combined site.
It turns out that from the view of concrete sheaves, a dominance is more-or-less a class of morphisms $\MCM$ on the site. We extract from this general setting a simple way of generating such a dominance, via a class $\MCM$ of `admissible' monomorphisms in the site $\bC$. For instance, in probabilistic programming, we would let $\MCM$ be generated by the Borel subsets, or for differentiable programming, $\MCM$ would be generated by the open subsets: these monomorphisms determine the notion of good domain for a partial function which is to be extended into the category of concrete sheaves (\S\ref{sec:cons-results}).
The synthetic domain theory foundation suggests elementary conditions that ensure that this class $\MCM$ combines well with the dominance of v-sets (Def.~\ref{def:class-of-admissible-monos}).

These conditions for a good notion of admissible monomorphism apply to all the examples from the literature we have considered so far. So we have a general framework for building models of functional programming languages with recursion and higher order functions: this is spelt out in \S\ref{sec:denot-semant-pcfv}. We emphasise the quality of these models with general soundness and adequacy theorems (Thms.~\ref{thm:wconc-adequate},~\ref{thm:sheaves-adequate}), connecting the interpretation in these models with operational semantics. 


%% file: concsheaves.tex
This paper is split in two parts. This first part is a self-contained exposition of $\omega$-concrete sheaves (\S\ref{sec:modelling-recursion}) as adequate models of our language (\S\ref{sec:lang}).
The second part explains why our constructions are canonical, by reference to synthetic domain theory (\S\ref{sec:beginning-of-sdt}).

\section{Categories of concrete sheaves}
\label{sec:concshf}
In this section we recall the definitions of concrete sites and concrete sheaves (\S\ref{sec:concr-sites-sheav}),
and examples of these constructions from the literature (\S\ref{sec:examplesconc}).

In brief, a \emph{concrete sheaf} is a set together with a collection of relations of different arities.
In this way, concrete sheaves are very close to logical relations models.
A \emph{site} specifies the number of these relations, what their arities are, and how the different relations should be connected. This is made precise by giving a category and a coverage on it. Later (\S\ref{sec:modelling-recursion}) we will also require a class of admissible monos in the site to capture notions of partiality. 

\subsection{Concrete sites and sheaves}\label{sec:concr-sites-sheav}
\newcommand{\defeq}{\stackrel{\text{def}}=}
\begin{definition}\label{def:omnibus-concrete}
  A \emph{concrete category} is a category $\bC$ with a terminal object $\star$ such that the functor $\bC(\star,-):{\bC\rightarrow\mathsf{Set}}$ is faithful.
  This means that morphisms $c \to d$ can identified with certain functions $|c| \to |d|$ where $\abs{c}\defeq \bC(\star,c)$ is the set of \emph{points}.
  In particular, $\abs{\star}$ is the singleton set and each map ${h:d\rightarrow c}$ is identified with a function ${\abs{h}:\abs{d}\rightarrow\abs{c}}$.

  A \emph{concrete site} $(\bC,J)$ is a small concrete category $\bC$ with an initial object $0$, together with a \emph{coverage} $J$, which specifies for each object $c$ 
a set $J(c)$ of families of maps with codomain~$c$. We call such a family $\{f_i:c_i\rightarrow c\}_{i\in I} \in J(c)$ a \emph{covering family} and say that it \emph{covers} $c$. The coverage must satisfy the following five axioms.
\begin{itemize}
\item [(C)] For every map $h:d\rightarrow c$ in $\bC$, if ${\{f_i:c_i\rightarrow c\}_{i\in I}}$ covers $c$, then there is a covering family ${\{g_j:d_j\rightarrow d\}_{j\in I'}}$ of~$d$ such that every $h\circ g_j$ factors through some $f_i$.
\item [($\star$)] If $\{f_i:c_i\rightarrow c\}_{i\in I}$ covers $c$, then $\bigcup_{i\in I}\mathsf{Im}(\abs{f_i})=\abs{c}$ (every covering family on $c$ contains all of its points).
\item [($0$)] The initial object $0$ is covered by the empty set.
\item [(M)] The identity is always covering: $\{ 1_c : c \to c \} \in J(c)$.
\item [(L)] If $\{f_i : c_i \to c\}_{i \in I} \in J(c)$ and $\{g_{ij} : c_{ij} \to c_i \}_{j \in J_i} \in J(c_i)$ for each $i$, then $\{ f_i \circ g_{ij} : c_{ij} \to c \}_{i \in I, j \in J_i} \in J(c)$.
\end{itemize}
\end{definition}

\begin{remark}\label{rem:concrete-site-apology}
  The more usual definition of `concrete site'~\cite{dubuc-concrete-quasitopoi,baez-hoffnung-smooth} would not require $\bC$ to have an initial object and would only require axioms (C) and ($\star$) for $J$.
  Since the same possible categories of concrete sheaves (Def.~\ref{def:concr-sheav}) can be presented, the restriction is inessential, but it does simplify our presentation especially regarding \Cref{def:class-of-admissible-monos}.
\end{remark}

\begin{definition}\label{def:concr-sheav}
  A \emph{concrete sheaf} $X$ on a concrete site $(\bC,J)$ is a set $\abs{X}$, together with, for each object $c\in\bC$, a set $R_X^c$ of functions of type $\abs{c}\rightarrow\abs{X}$, such that:
\begin{itemize}
\item Each $R^c_X$ contains all the constant functions.
\item For any map $h:d\rightarrow c \in \bC$, and any $g\in R^c_X$, the composite function $g\circ\abs{h}:\abs{d}\rightarrow\abs{X}$ is in $R^d_X$.
\item For each function $g:\abs{c}\rightarrow\abs{X}$ and each covering family $\{f_i:c_i\rightarrow c\}_{i\in I}$, if each $g\circ\abs{f_i}\in R^{c_i}_X$, then $g:\abs{c}\rightarrow\abs{X} \in R^c_X$.
\end{itemize}
A morphism $\alpha:X\rightarrow Y$ between concrete sheaves is a function $\alpha:\abs{X}\rightarrow\abs{Y}$ that preserves the structure, namely if $g\in R^c_X$, then $\alpha\circ g \in R^c_Y$.

The concrete sheaves on a concrete site $(\bC,J)$ form a category $\conc(\bC,J)$ which is cartesian closed and has coproducts, so it can interpret simply-typed lambda-calculus with sums.
\end{definition}
\subsection{Examples}
\label{sec:examplesconc}
\begin{example}[Reflexive relations]\label{ex:reflrel}
  Consider the category whose objects are sets $\abs X$ equipped with a binary relation
  $R_X\subseteq \abs X^2$ such that $(x,x)\in R_X$ for all $x$, and where the morphisms are functions $\abs X\to \abs Y$ that preserve the relation ($(x,x')\in R_X\implies (f(x),f(x'))\in R_Y$).
  This is a model that might be used in a simple logical relations argument (e.g.~\cite{plotkin-lambda-definability-and-logical-relations}). 
  This category is a category of concrete sheaves.
  For the site, take the category generated by three objects $0,\star,2$ and two morphisms $\star\rightrightarrows 2$ all such that $0$ and $\star$ are initial and terminal respectively.
  Then $\abs 2$ has two elements, and we can regard~$R_X^2\subseteq [\abs 2\to \abs X]\cong \abs X^2$ as a binary relation.
  The coverage $J$ is the trivial one, where $2$ and $\star$ are covered by identities and $0$ by the empty set.
\end{example}
\begin{example}[Probability and measure~\cite{qbs,DBLP:journals/pacmpl/VakarKS19}]\label{ex:qbs}
  Quasi-Borel spaces are a setting that incorporates probability theory and higher order functions. A quasi-Borel space is a set $X$ together with a set $R_X^\bR\subseteq [\bR\to X]$ of admissible random elements in $X$, satisfying some conditions. These are quite widely used (e.g.~\cite{DBLP:journals/pacmpl/ScibiorKVSYCOMH18,DBLP:journals/pacmpl/SatoABGGH19,DBLP:journals/pacmpl/AguirreBGGKS21,DBLP:journals/pacmpl/LewCSCM20}).
  As is well known, the category of quasi-Borel spaces $\mathbf{Qbs}$ can be regarded as the category of concrete sheaves on a site $(\mathbf{Sbs}, J_\mathbf{Sbs})$.
  Here the category $\mathbf{Sbs}$ has as objects the Borel subsets of $\bR$ with morphisms all the measurable functions between these objects. The coverage $J_\mathbf{Sbs}(U)$ contains the countable sets of inclusion functions $\{U_i\hookrightarrow U\}_{i\in I}$ such that $U=\bigcup_{i\in I}U_i$ and the $U_i$'s are disjoint.
\end{example}

\begin{example}[Smoothness~\cite{huot-staton-vakar,DBLP:journals/corr/abs-2007-05282}]\label{ex:smooth}
  Diffeological spaces are a setting that incorporates smoothness with higher order functions~\cite{diffeology-book}. A diffeological space is a set $X$ together with a set $R_X^U\subseteq [U\to X]$ of admissible plots from each open subspace $U$ of a Euclidean space, satisfying some conditions.
  As is well known, the category of diffeological spaces $\mathbf{Diff}$ can be regarded as the category of concrete sheaves on a site $(\mathbf{Cart},J_\mathbf{Cart})$~\cite{smootheology,baez-hoffnung-smooth}. Here, the objects of $\mathbf{Cart}$ are the open subsets $U\subseteq \bR^n$ for any $n\in\bN$, and morphisms are smooth maps. The coverage $J_\mathbf{Cart}(U)$ contains the countable sets of inclusion functions $\{U_i\hookrightarrow U\}_{i\in I}$ such that $U=\bigcup_{i\in I}U_i$.
\end{example}

\begin{example}[Piecewise smoothness and $\mathbf{PAP}$~\cite{DBLP:journals/corr/abs-2111-15456}]
  \label{ex:pap}
  Recently a variation on diffeological spaces has been proposed that allows a controlled degree of non-smoothness.
  The idea is to consider sets of plots $R_X^U\subseteq [U\to X]$ that are indexed by `c-analytic' sets, rather than Euclidean open sets: these are sets $U\subseteq \bR^n$ for some $n$ that are countable unions of analytic subsets.
  The resulting `PAP-sets' can be regarded as the category of sheaves on the site
  $(\mathbf{PAP},J_\mathbf{PAP})$, where the objects are c-analytic subsets, and morphisms are $\mathbf{PAP}$ functions between them \cite{DBLP:conf/nips/0001YRY20,DBLP:conf/aistats/ZhouGKRYW19}. The coverage $J_\mathbf{PAP}(U)$ contains countable sets of inclusion functions $\{A_i\hookrightarrow U\}_{i\in I}$ where $(A_i)_i$ are disjoint c-analytic sets such that $U=\bigcup_{i\in I}A_i$.
  The category of concrete sheaves on $(\mathbf{PAP},J_\mathbf{PAP})$ models the fragment without recursion of the differentiable language from~\cite{DBLP:journals/corr/abs-2111-15456}.
\end{example}

\begin{example}[Topological examples]
  \label{ex:top}
  Arguably the earliest examples of concrete sheaves arose from finding convenient categories of topological spaces.
  For example, a `subsequential space' is a set $\abs X$ together with a set of convergent sequences in $X$ equipped with their limits, i.e.~a set of functions $R_{\abs X}\subseteq [\mathbb{N}\cup\{\infty\}\to \abs X]$ satisfying some conditions~\cite{topological-topos}. A `sequentially continuous function' is a function that preserves this sequence structure.
  As discussed in~\cite{topological-topos}, subsequential spaces can be viewed as concrete sheaves on the site
  whose objects are $0$, $\star$ and $\bN\cup\{\infty\}$, and whose morphisms are continuous functions. 

  There are several related categories. For example, \emph{C-spaces} \cite{escardo-xu} arise in a similar way but  replacing $(\bN\cup\{\infty\})$ with the Cantor space $2^\bN$. In~\Cref{sec:vset} we will replace $(\bN\cup\{\infty\})$ with the vertical natural numbers equipped with the Scott topology, following~\cite{fiore-rosolini-h}.
\end{example}
\begin{example}[Quantum sets]
  The construction of concrete sheaves can be considered whenever we have a concrete category modelling some computational phenomena.
  For example, to model quantum computation, consider the category whose objects are natural numbers $n$ regarded as sets $\mathcal{DM}_n$ of density matrices, i.e.~$n\times n$ complex matrices that are positive, semidefinite and with trace~$1$.
  The morphisms are quantum channels, i.e.~completely positive trace-preserving maps~\cite{nielsen-chuang}.
  This is a concrete category, and $0$ is initial and $1$ is terminal.
  We can thus consider concrete sheaves on this category with the trivial coverage (i.e.~concrete presheaves).
  These are sets $\abs X$ equipped with sets of maps $R^n_X\subseteq [\mathcal{DM}_n\to \abs X]$,
  regarded as the admissible quantum channels into $\abs X$.
  This example is a concrete variation on the presheaf models of quantum computation considered in e.g.~\cite{mss-presheaf-quantum,lmz-quantum}.
\end{example}
There are many other examples of categories of concrete sheaves across computer science and mathematics (e.g.~\cite{dubuc-concrete-quasitopoi,ehrhard-concrete,rosolini-streicher-concrete}). 
Here we have focused on examples for which the methods in the following section are useful in modelling recursion in programming language semantics.


%% file: cpo-with-functions.tex
\section{Concrete sheaves with recursion}
\label{sec:modelling-recursion}

In this section we introduce the new general idea: to get a model of call-by-value PCF we extend our attention to concrete sheaves with an $\omega$cpo structure, and discover a well-behaved notion of partiality and lifting via classes of admissible monos. 

\subsection{$\omega$-Concrete sheaves}
Recall that an $\omega$cpo is a partially ordered set closed under least upper bounds of countable chains. A continuous function between $\omega$cpo's is a monotone function that preserves least upper bounds (e.g.~\cite{winskel-semantics}).
\begin{definition}
An \emph{$\omega$-concrete sheaf} on a site $(\bC,J)$ is a concrete sheaf $X$ together with an ordering $\leq_X$ on $\abs{X}$ that gives $\abs{X}$ the structure of an $\omega$cpo, such that each $R^c_X$ is closed under pointwise suprema of countable chains with respect to the pointwise ordering.

A morphism $\alpha:X\rightarrow Y$ of $\omega$-concrete sheaves is a continuous function between $\omega$cpo's, $\alpha:\abs{X}\rightarrow\abs{Y}$, that is also a morphism of concrete sheaves. $\omega$-concrete sheaves form a category $\omega\mathsf{Conc}(\bC,J)$, which is a cartesian closed category with binary coproducts. 
\end{definition}

\subsection{Admissible monos, a lifting monad, \& partiality}\label{sec:admiss-monos-lift}
To model recursion we first need to define a (strong) lifting monad $L$ on $\omega\mathsf{Conc}(\bC,J)$.
Recall that a monad~\cite{moggi-metalanguage} is a triple $(L,\{\eta_X:X\rightarrow LX\}_X,\{\mu_X:LLX \rightarrow LX\}_X)$ satisfying some identity and associativity equations. Furthermore, $L$ is strong if there is a family of maps $\{\mathsf{st}_{X,Y}:X\times LY\rightarrow L(X\times Y)\}_{X,Y}$ satisfying some conditions; in a concrete category, if the strength exists, it is determined uniquely by $L$ and the cartesian structure of the category~\cite[Prop.~3.4]{moggi-metalanguage}.

For an $\omega$-concrete sheaf $X$, we can define the lifting monad $L$ to have underlying $\omega$cpo $\abs{LX}=\abs{X}\uplus\{\bot\}$, just like in the case of the lifting monad in the $\omega$cpo-model of call-by-value PCF. However, it is not immediately apparent how to define $R^c_{LX}\subseteq [\abs{c}\rightarrow\abs{X}\uplus\{\bot\}]$, there are many choices. For this reason we parametrize the definition of the lifting monad by a class $\MCM$ of monomorphisms from the site $(\bC,J)$, which we call \emph{admissible monos}. The intuition is that the admissible monos $c'\rightarrowtail c$ are the possible domains of partial functions $\abs{c}\rightarrow\abs{X}$ from $R^c_{LX}$.

Recall that, in any category, monos with the same codomain are preordered: if $m : d \rightarrowtail c, m' : d' \rightarrowtail c$ then $m \leq m'$ iff there exists $f : d \to d'$ with $m' \circ f = m$.
We write $\sub(c)$ for the poset quotient of the set of monos with codomain $c$.
For any class $\MCM$ of monos in $\bC$, we write $\sub_\MCM(c)$ for the poset of $\MCM$-subobjects, i.e.\ the full subposet of $\sub(c)$ whose elements have representatives in $\MCM$.
If, moreover, we suppose that all pullbacks of maps in $\MCM$ exist (along any map in $\bC$) and are again in $\MCM$, then $\sub_\MCM$ can be viewed as a functor $\bC^\op \to \mathsf{Poset}$.
If $(\bC,J)$ is a concrete site, then there is a natural transformation $\alpha$:
\begin{displaymath}
  \alpha_c:\sub_\MCM(c) \to \sets(|c|,\{0,1\})
\end{displaymath}
where $m : c' \rightarrowtail c$ is taken to the function sending $p : \star \to c$ to $1$ if $p$ factors through $m$ and $0$ otherwise.
Naturality means that pullback along $f : c' \to c$ becomes precomposition by $|f| : |c'| \to |c|$, and indeed each component of the transformation is a monotone map for the obvious pointwise ordering on $\sets(|c|,\{0,1\})$.

\begin{definition}\label{def:class-of-admissible-monos}
  A class $\MCM$ of \emph{admissible monos} from $(\bC,J)$ consists of, for each object $c\in\bC$, a set of monos $\MCM(c)$ with codomain $c$ satisfying the following conditions.
  \begin{enumerate}
  \item For all $c \in \bC$, $0\xrightarrow{!}c \in \MCM(c)$.
  \item $\MCM$ contains all isomorphisms.
  \item $\MCM$ is closed under composition: if $f:c''\rightarrowtail c' \in \MCM(c')$ and $g:c'\rightarrowtail c\in\MCM(c)$, then $g\circ f\in \MCM(c)$.
  \item All pullbacks of $\MCM$-maps exist and are again in $\MCM$.
    (This makes $\sub_\MCM$ a functor $\bC^\op \to \mathsf{Poset}$.)
  \item For each $c$, the function $\alpha_c:\sub_\MCM(c) \to \sets(|c|,\{0,1\})$ is componentwise injective and order-reflecting, and the image of $\sub_\MCM(c)$ is closed under suprema of $\omega$-chains.
  \item Given an increasing chain in $\MCM(c)$, $(c_n\rightarrowtail c)_{n\in\mathbb{N}}$, denote its least upper bound by $c_\infty\rightarrowtail c$.
    Then the closure under precomposition (with any morphism) of the set $\{ c_n\rightarrowtail c_\infty\}_{n\in\mathbb{N}}$ contains a covering family of $c_\infty$.
  \end{enumerate}
\end{definition}
To spell this definition out a little: when $\MCM$ is a class of admissible monos, $\MCM$-subobjects $m : c' \rightarrowtail c$ of $c \in \bC$ are determined by the induced inclusions of sets $\im(\abs m) \subseteq \abs c$, and the order relation is given by $m \leq m'$ iff $\im(\abs m) \subseteq \im(\abs{m'})$.
In particular, $\sub_\MCM(\star)$ has at most two elements, corresponding to the two subsets of the one-element set $\abs{\star}$.
Since in a concrete site $0$ has an empty cover and $\star$ does not, they are not isomorphic and hence we actually see that $|\sub_\MCM(\star)| = 2$.
Moreover, the suprema of an $\omega$-chain $\{m_0 \leq m_1 \leq \ldots\} \in \sub_\MCM(c)$ exists and is given by the unique $m_\infty \in \sub_\MCM(c)$ such that $\im(\abs{m_\infty}) = \bigcup_n \im(\abs{m_n})$.

\begin{definition}
We can define the (strong) lifting monad $L_\MCM$ associated to the class of admissible monos $\MCM$ as:
\begin{align*}
&  \abs{L_\MCM X}=\abs{X}\uplus\{\bot\} \\
  &\forall x\in\abs{X}.\, \bot\leq_{L_\MCM X} x,\quad \forall x,x'\in\abs{X}. x\leq_{L_\MCM X} x' \text{ iff } x\leq_X x' \\
  &R^c_{L_\MCM X}=\big\{g:\abs{c}\rightarrow \abs{X}\uplus\{\bot\} ~\big|~ \exists c'\rightarrowtail c \in \MCM(c) \text{ s.t.}\\
  &\hspace{2.5cm}g^{-1}(\abs{X})=\mathsf{Im}(\abs{c'}) \text{ and } g\vert_{\mathsf{Im}(\abs{c'})} \in R^{c'}_X \big\} \\
  \intertext{The strong monad structure is exactly the same as the `maybe' monad on $\Set$ \cite{moggi-metalanguage}, which one can check preserves all the structure. Here $\abs{\eta_X}:\abs{X}\rightarrow \abs{L_\MCM X}$,
  $\abs{\mu_X}:\abs{L_\MCM L_\MCM X} \rightarrow \abs{L_\MCM X}$,
  $\abs{\mathsf{st}_{X,Y}}:\abs{X}\times \abs {L_\MCM Y} \rightarrow \abs{L_\MCM (X\times Y)}$:}
  &\abs{\eta_X}(x)=x \qquad 
  \abs{\mu_X}(x)=x, \qquad
    \abs{\mu_X}(\bot_1)=    \abs{\mu_X}(\bot_2)=\bot,\\
  &\abs{\mathsf{st}_{X,Y}}(x,y)=(x,y),\qquad
  \abs{\mathsf{st}_{X,Y}}(x,\bot)=\bot\text.
\end{align*}
\end{definition}

The lifting monad induces a notion of partial map.
Recall that to give a total function $\abs X\to \abs Y\uplus \{\bot\}$ is to give a partial function
$\abs X\rightharpoonup\abs Y$.
\begin{proposition}\label{prop:partialmap}
  A partial function $f:\abs X\rightharpoonup\abs Y$ between $\omega$-concrete sheaves
  corresponds to a morphism $X\to L_\MCM Y$ if and only if it is continuous, its domain is Scott-open (i.e. the characteristic function of the domain into the preorder $\{0\leq 1\}$ is continuous), and for any $g:\abs{c}\rightarrow\abs{X}\in R^c_X$, the domain of the partial function $f\circ\abs{g}:\abs{c}\rightharpoonup\abs{Y}$ is determined by an $\MCM$-subobject $c'\rightarrowtail c$, and $(f\circ\abs{g})|_{\mathsf{Im}(\abs{c'})}\in R^{c'}_Y$.
\end{proposition}
\begin{proof}[Proof note] By expanding the definitions. \end{proof}
\begin{proposition}\label{prop:wconc-fp}
  There is a fixed point combinator, a morphism  $((L_\MCM Y)^X\Rightarrow(L_\MCM Y)^X)\rightarrow (L_\MCM Y)^X$, in $\omega\conc(\bC,J)$.
\end{proposition}
\begin{proof}[Proof notes]
  A candidate fixed point can be constructed just like in the $\omega$cpo model of \pcfv, using the $\omega$cpo structure of $\abs{X}$ and $\abs{LY}$ and Tarski's fixed point theorem. 
  It then remains to show this candidate fixed point preserves the structure of concrete sheaves; this is where the last property in the definition of $\MCM$ is needed.
See also~\Cref{sec:sheaf-model-pcfv}.
\end{proof}


\subsection{Examples}
\label{sec:exampleswconc}
\begin{example}[Probability and measure ctd.]
  The category of $\omega$-concrete sheaves on $(\mathbf{Sbs}, J_\mathbf{Sbs})$ is equivalent to $\omega\mathbf{Qbs}$~\cite{DBLP:journals/pacmpl/VakarKS19}.
  If we choose the admissible monomorphisms such that $\MCM_\mathbf{Sbs}(U)$ contains all the monos with codomain $U$, then the induced lifting monad is the one  used in~\cite{DBLP:journals/pacmpl/VakarKS19} to model recursion.
\end{example}

\begin{example}[Smoothness ctd.]
  The category of $\omega$-concrete sheaves on $(\mathbf{Cart},J_\mathbf{Cart})$ is equivalent to $\omega\mathbf{Diff}$~\cite{DBLP:journals/corr/abs-2007-05282}.
  Consider the class of admissible monos such that $\MCM_\mathbf{Cart}(U)$ contains the open inclusion maps into $U$, then the induced lifting monad is the one used to model recursion in~\cite{DBLP:journals/corr/abs-2007-05282}.
\end{example}

\begin{example}[Piecewise smoothness and $\mathbf{PAP}$ ctd.]
  The category of $\omega$-concrete sheaves on $(\mathbf{PAP},J_\mathbf{PAP})$ is equivalent to the category used in~\cite{DBLP:journals/corr/abs-2111-15456} to model a higher-order differentiable language with recursion.
  Choose the class of admissible monos $\MCM_\mathbf{PAP}$ to contain at $U$ the c-analytic subsets $U'\hookrightarrow U$. The induced lifting monad gives the same notion of partial map as the one used in~\cite{DBLP:journals/corr/abs-2111-15456}.  
\end{example}

\begin{example}[Fully abstract models of PCF~\cite{matache-moss-staton-fscd-2021,DBLP:journals/iandc/RieckeS02a}]
  In~\cite{matache-moss-staton-fscd-2021} we present a fully abstract sheaf model $\MCG$, on a concrete site, for call-by-value PCF (explained further in~\Cref{ex:fscd}).
  The lifting monad we use there is obtained from a dominance which is actually equivalent to an class of admissible monos.
  In fact, the interpretation of PCF lies in the subcategory of $\omega$-concrete sheaves of $\MCG$.
  This category of $\omega$-concrete sheaves is very similar to the logical relations (fully abstract) FPC  model proposed by Riecke and Sandholm~\cite{DBLP:journals/iandc/RieckeS02a}, where an object is roughly a cpo equipped with  relations of varying arity.
  \end{example}
  We can also consider admissible monos on other sites. For example, Example~\ref{ex:top} suggests a candidate semantic model for the local continuity in~\cite[\S6]{DBLP:conf/esop/BartheCLG20}.

  Here and in Section~\ref{sec:examplesconc} we have focused on the models of these kinds of phenomena that are based on $\omega$-concrete sheaves. Of course, there are other ways to give semantic models of higher order recursion, including for probabilistic programming~\cite{DBLP:journals/pacmpl/EhrhardPT18,DBLP:conf/lics/Crubille18,DBLP:journals/pacmpl/DahlqvistK20,huang_morrisett_spitters_2020,DBLP:journals/corr/abs-2106-16190,DBLP:conf/lics/AmorimKMPR21}, differentiable programming~\cite{DBLP:journals/pacmpl/BrunelMP20}, and
  full abstraction~\cite{DBLP:journals/iandc/HylandO00,DBLP:journals/iandc/AbramskyJM00,saville-kammar-katsumata-fullabs}.

\subsection{Conservativity results}\label{sec:cons-results}

One major application of $\omega$-concrete sheaves is in giving conservativity results for a programming language over a particular class of functions. For example, suppose we write programs in a language with higher order recursion and a type $\treal$ of real numbers. If all the primitive functions are continuous, does that mean that the definable functions $\treal\to\treal$ are all continuous?

In general, suppose we have some set~$X$ and a class $\mathscr{C}$ of operations over it. If we write programs over $X$ in a language with higher order recursion, are the definable functions $X\to X$ in the class $\mathscr{C}$? 
In a language with recursion, the definable functions need not terminate and so might be partial. Thus more precisely, we should investigate the definable partial functions $X\to X$.
We would characterize these partial functions, and their domains of definition. 
For example, if all the primitive operations are continuous, then we might prove that the definable functions $\treal\to \treal$ are partial continuous functions whose domain is an open set. 
The theory of $\omega$-concrete sheaves is a good setting for this.

\begin{definition}A concrete site $(\bC,J)$ is \emph{subcanonical} if for all ${c\in\bC}$ the relations
  $\Big(R^d_c=\big\{\abs f\colon \abs d\to \abs c~\big|~f\colon d\to c\big\}\Big)_{d\in\bC}$
  form a concrete sheaf over $\abs c$.
\end{definition}
All the examples in Section~\ref{sec:examplesconc} are subcanonical. (The site for full abstraction in Ex.~\ref{ex:fscd} is not subcanonical, though.)

For a subcanonical site, we have a functor $Y:\bC\to \omegaConc(\bC,J)$ given by
$Y(c)=c$ with the discrete order, and with $Y(f)=\abs f$. (This is the Yoneda embedding, which is particularly simple for concrete sheaves.)

For a subcanonical site, we can also define a category of partial maps.
Recall the following quite general construction.

\begin{definition}\label{def:m-partial-maps}
  Let $\bC$ be any category and $\MCM$ any class of monos containing the isomorphisms, closed under composition, and all of whose pullbacks exist and are again in $\MCM$.
  Then $\pC_\MCM$, the \emph{category of $\MCM$-partial maps in $\bC$}, has the same objects as $\bC$ but morphisms $c \to b$ are equivalence classes of pairs $({m : d \rightarrowtail c},$  ${ f : d \to b})$ with $m \in \MCM$, where $({m : d \rightarrowtail c}, {f\colon d \to b})$ is equivalent to $(m' : d' \rightarrowtail c, f' : d' \to b)$ iff there exists an isomorphism $t : d \to d'$ with $m' \circ t = m$ and $f' \circ t = f$.
  This really is a category: one uses the pullback-stability of $\MCM$ to compose partial maps.
\end{definition}

For our concrete site $(\bC,J)$, we can describe partial maps $\pC_\MCM(c,d)$ as partial functions $f : |c| \rightharpoonup |d|$ such that there is a monomorphism $m : c' \rightarrowtail c$ in $\MCM$ with $\im(|m|) = \dom f$ and a morphism $h : c' \to d$ in $\bC$ such that $|h| = f \circ |m|$.

We relate this category $\pC_\MCM$ of partial maps to the category of partial maps between $\omega$-concrete sheaves, which is the Kleisli category of $L$. 
There is a functor $Z:\pC_\MCM\to \mathbf{Kl}(L)$, given by $Z(c)=c$, and with $Z(f)(x)=f(x)$ if $x\in \mathsf{dom}(f)$, and $Z(f)(x)=\bot$ if $x\not\in \mathsf{dom}(f)$. 
\begin{theorem}\label{thm:conserv}
  If $(\bC,J)$ is a subcanonical concrete site with an admissible class of monos $\MCM$, then the functors
  $Y:\bC\to \omegaConc(\bC,J)$ and $Z:\pC_\MCM\to \mathbf{Kl}(L_\MCM)$
  are full and faithful.
\end{theorem}
 Thus the morphisms $\bC(c,d)$ are in bijection with the morphisms
$\omegaConc(\bC,J)(c,d)$, and the partial maps $\pC_\MCM(c,d)$ are in bijection with the
 Kleisli maps $\mathbf{Kl}(L)(c,d)$.
So $\omegaConc(\bC,J)$ has powerful structure for interpreting recursion (Prop.~\ref{prop:wconc-fp}) and higher order functions, but is conservative in that it agrees with $\bC$ on morphisms and partial maps. 
 \begin{proof}[Proof notes for Thm.~\ref{thm:conserv}]This is a Yoneda argument, but can also be checked directly by expanding the definitions, using Prop.~\ref{prop:partialmap}. \end{proof}
Although this conservativity result is new, it is reminiscent of earlier representation results for partiality~\cite{mulry-partial-map-classifiers-and-partial-cccs}, axiomatic domain theory~\cite{fiore-unpub},
and computational effects~\cite{DBLP:conf/tlca/Power03}.

 This theorem generalizes some known useful facts, such as, to give a partial measurable function $\bR\to \bR$ with Borel domain is to give a morphism $\bR\to L(\bR)$ of $\omega$-quasi-Borel spaces~\cite{DBLP:journals/pacmpl/VakarKS19}; to give a partial smooth function $\bR\to \bR$ with open domain
is to give a morphism $\bR\to L(\bR)$ between $\omega$-diffeological spaces~\cite{DBLP:journals/corr/abs-2007-05282}.
On top of this, the functor $Y$ always preserves limits that exist and the sheaf condition can be understood as saying that $Y$ preserves certain colimits. For example, in $\omega$-quasi-Borel spaces, the coproduct $\bN=1{+}1{+}1{+}\dots$ is such that the morphisms $\bR\to \bN$ are the Borel partitions of~$\bR$. 


%% file: language.tex
\section{A higher-order language with recursion}
\label{sec:lang}
To illustrate the constructions in the previous section, we discuss \pcfv, a call-by-value simply typed lambda calculus with recursion. We give an operational semantics (\S\ref{sec:opsem}) and a denotational semantics in $\omega$-concrete sheaves (\S\ref{sec:denot-semant-pcfv}), which we show to be adequate (Thm.~\ref{thm:wconc-adequate}). We allow the language to be extended with new type constants (such as $\treal$) and functions, inspired by the ability of concrete sheaves to provide new models of higher-order recursion that incorporate other constructions (\S\ref{sec:cons-results}). 

\subsection{PCF and its operational semantics}
\label{sec:opsem}
In our formulation of \pcfv, there is a syntactic distinction between values and computations, which means the calculus is fine-grained~\cite{levy-power-thielecke}. The grammars of types, values and computations are:
\begin{align*}
& \tau,\tau' \Coloneqq \nat \mid \tau\rightarrow\tau'\qquad
  v,w \Coloneqq x \mid \vzero \mid \suc{v} \mid  \rec{f}{x}{t}\\
&  t  \Coloneqq v\ w \mid \ret{v} \mid  \letin{x}{t}{t'} \mid  \casenat{v}{t}{x}{t'}
\end{align*}
The value $(\rec{f}{x}{t})$ is a recursive function definition, which can be thought of as $f(x)=t$.
When $f$ does not appear in $t$, we can write $\lbd{x}{t}$.
The computation $\letin{x}{t}{t'}$ sequences computations $t$ and $t'$.

There are two typing relations, one for values, $\vterm$, and one for computations, $\cterm$.
\begin{gather*}
 \inferrule{\Gamma,\, f:\tau\rightarrow\tau',\, x:\tau \cterm t:\tau'}{\Gamma \vterm \rec{f}{x}{t} : \tau\rightarrow \tau'} \quad
 \inferrule{\Gamma \vterm v :\tau\rightarrow \tau' \quad\Gamma \vterm w :\tau}{\Gamma \cterm v\ w : \tau'}\\
  \inferrule{\Gamma \vterm v :\tau}{\Gamma \cterm \ret{v} : \tau} \quad   \inferrule{\Gamma \cterm t :\tau \\ \Gamma,x:\tau \cterm t :\tau'}{\Gamma \cterm \letin{x}{t}{t'} : \tau'}
\\
\inferrule{\phantom{\Gamma'\vterm}-\phantom{\Gamma'\vterm}}{\Gamma,x:\tau,\Gamma'\vterm x:\tau} \quad
  \inferrule{\phantom{\Gamma'\vterm}-\phantom{\Gamma'\vterm}}{\Gamma \vterm \vzero :\nat} \quad
  \inferrule{\Gamma\vterm v: \nat}{\Gamma \vterm \suc{v}: \nat}
   \end{gather*}

\begin{gather*}
  \inferrule{\Gamma \vterm v:\nat \\ \Gamma \cterm t :\tau \\ \Gamma,x:\nat \cterm t':\tau}{\Gamma \cterm \casenat{v}{t}{x}{t'} : \tau} \\
\end{gather*}
The big-step operational semantics of \pcfv\ is a relation between closed computations and closed values. It is the least relation closed under the rules below:
\newcommand{\Eval}[1]{\Downarrow}
{\small\begin{gather*}
  \inferrule{ }{\ret{v} \Eval\tau v} \quad
  \inferrule{t[(\rec{f}{x}{t})/f,\,v/x] \Eval{\tau} w}{(\rec{f}{x}{t})\ v \Eval{\tau} w} \quad 
  \inferrule{t \Eval{\tau} v \\ t'[v/x] \Eval\tau w}{\letin{x}{t}{t'} \Eval\tau w} \\
  \inferrule{t \Eval\tau w}{\casenat{\vzero}{t}{x}{t'} \Eval\tau w} \quad
  \inferrule{t'[v/x] \Eval\tau w}{\casenat{\suc{v}}{t}{x}{t'} \Eval\tau w}
\end{gather*}}

By induction on the structure of typing derivations, if $\cterm t:\tau$ and $t\Eval \tau v$ then $\vterm v:\tau$. 

This calculus is chosen to be simple enough to illustrate the key ideas. We can further add sum and product types, as outlined in~\Cref{app:typing:rules}.

\newcommand{\vals}[1]{\mathrm{Val}_{#1}}
We can extend our calculus further: let $\alpha,\beta$ denote new type constants which we want to add, or $\nat$ (i.e.~ground types).
We can then add term constant $f:\alpha\to \beta$.
For example, with an eye to several of the examples in~\Cref{sec:examplesconc}, we might add a new type constant $\treal$ for real numbers. Depending on the application, we could add term constants such as $\sin,\arctan:\treal\to\treal$,
or a non-smooth function such as $\max(0,-):\treal\to\treal$, or a discontinuous function such as rounding $\treal\to \nat$.

To extend the operational semantics, we suppose that every new type constant $\alpha$ is associated with a set $\vals \alpha$ of values. For example, we would likely put $\vals\treal=\bR$. This extends the basic language which has $\vals \nat\cong\bN$. We then require that every new term constant $f:\alpha\to \beta$ is associated with a partial function $\vals\alpha\to\vals\beta$.
Then we add all the elements of these sets as values, and implement a straightforward operational semantics:
\[
  \inferrule{-}{\vterm f:\alpha\to\beta}
  \quad
  \inferrule{-}{\vterm c:\alpha}{(c\in \vals\alpha)}
  \quad
  \inferrule{-}{f\,v\Eval\beta w}( f(v)=w)
\]



%% file: adequacy.tex
\subsection{Denotational semantics for \pcfv}\label{sec:denot-semant-pcfv}

Given a concrete site with an admissible class of monos, $(\bC,J,\MCM)$, we can interpret \pcfv\ types using the structure of the category $\omega\mathsf{Conc}(\bC,J)$ of $\omega$-concrete sheaves as: 
\begin{gather*}
\den\nat = {\textstyle \sum_0^\infty 1 = 1 + 1 + \ldots} \quad  \den{\tau\rightarrow \tau'}=\den{\tau}\Rightarrow L_\MCM\den{\tau'}
\end{gather*}
whose explicit description is given in~\Cref{fig:types-wconc}.
A value $\Gamma\vterm v:\tau$ is interpreted as a map $\den\Gamma\rightarrow\den\tau$ and a computation $\Gamma\cterm t:\tau$ as a map $\den\Gamma\rightarrow L_\MCM \den\tau$. The interpretation of both values and computations is standard~\cite{moggi-metalanguage}; to interpret $(\rec{f}{x}{t})$ we use the fixed point from~\Cref{prop:wconc-fp}.

\begin{figure*}
  \begin{equation*}
  \begin{aligned}
    \abs{\den\nat} &=\bN \text{, with the discrete order} \qquad
    \abs{\den{\tau\rightarrow \tau'}} = \omega\conc(\den\tau,\,  L_\MCM\den{\tau'}), \text{ with the pointwise order} \\
    R^c_{\den\nat} &= \big\{f:\abs{c}\rightarrow\bN \midd \exists\, \{g_i:c_i\rightarrow c\}_{i\in I}\in J(c) \text{ s.t. each } f\circ g_i \text{ is constant} \big\}\\
    R^c_{\den{\tau\rightarrow \tau'}} &= \big\{f:\abs{c}\rightarrow \omega\conc(\den\tau,\,  L_\MCM\den{\tau'}) \midd \forall h:d\rightarrow c\in\bC,\ \forall g:\abs{d}\rightarrow\abs{\den\tau} \in R^d_{\den\tau}.\ \lambda x\in\abs{d}.\big(f(h(x))\ g(x)\big)\in R^d_{L_\MCM \den{\tau'}}\big\}
  \end{aligned}
  \end{equation*}
  \caption{Interpretation of types in $\omega\conc(\bC,J)$}
  \label{fig:types-wconc}
\end{figure*}

 We can extend the interpretation to the setting with new type constants (such as $\treal$) and term constants. Given the set of values $\vals\alpha$ of a type constants $\alpha$, we must equip $\vals\alpha$ with the structure of an $\omega$-concrete sheaf ${\den\alpha}$.
The interpretation works as long as the function corresponding to each term constant $f:\vals\alpha\to\vals\beta$ is in fact a partial morphism $\den\alpha\to\den\beta$ of concrete sheaves (i.e.~$\den{f}:\den\alpha\rightarrow L_\MCM \den\beta$, via Prop.~\ref{prop:partialmap}). Several of the examples in Section~\ref{sec:examplesconc} admit different structures for $\treal$. In each case, the underlying set is $\bR$, but we can equip this with the structure of all Borel morphisms (Ex.~\ref{ex:qbs}, admitting almost all term constants of interest), all smooth plots  (Ex.~\ref{ex:smooth}, forbidding functions like $\max(0,-)$), plots that are piecewise smooth under analytic partition (Ex.~\ref{ex:pap}), or sequentially continuous functions (Ex.~\ref{ex:top}, forbidding rounding $\treal\to\nat$). 
More generally, if one is faced with a new class of type and term constants, the methods of Section~\ref{sec:cons-results} could be used to generate a site for suitable $\omega$-concrete sheaves. 

\begin{theorem}\label{thm:wconc-adequate}
  The model (in $\omega\conc(\bC,J)$) of \pcfv\ presented by $(\bC,J,\MCM)$ is sound and adequate.
  That is, for closed terms $\cterm t: \tau$ and $\vterm v: \tau$:
    \begin{itemize}
  \item Soundness: $t\Downarrow v \implies \den t = \den{\ret{v}} \in L_\MCM\den\tau$.
  \item Adequacy: if $\tau$ is a ground type ($\nat$ or a type constant $\alpha$), then $\den t = \den{\ret{v}} \implies t \Downarrow v$. 
  \end{itemize}
\end{theorem}
The proof of this theorem is deferred to~\Cref{sec:sheaf-model-pcfv}. It still holds when we add product and sum types, as in~\Cref{app:typing:rules}. 


%% file: dominance.tex
\label{sec:beginning-of-sdt}
We recall how dominances give rise to partiality and lifting in general (\S\ref{sec:dominance}),
before specializing the constructions to categories of sheaves (\S\ref{sec:sheaves}) to connect
dominances to classes of pre-admissible monos (\S\ref{sec:preadmissible}). 
\subsection{Dominances and lifting in general}
\label{sec:dominance}

The construction of the lifting monad in~\Cref{sec:admiss-monos-lift} is actually a special case of a more general construction.
We recall the following definition originally from \cite{rosolini-phd} but given a more general formulation as in \cite{fiore-plotkin-an-extension-of-models-of-adt-to-models-of-sdt}.

\begin{definition}
  Let $\MCE$ be a category with a terminal object~$1$.
  A \emph{dominance} is a monomorphism $\top : 1 \rightarrowtail \Delta$ in $\MCE$ such that
  \begin{enumerate}
  \item all pullbacks of $\top$ exist, and
  \item for all $A \in \MCE$ the function $\MCE(A,\Delta) \to \sub(A)$ given by pullback along $\top$ is an injection.
  \end{enumerate}
\end{definition}

\begin{remark}
  The traditional setting for a dominance is a topos, wherein the first condition in the definition above is redundant.
  Moreover, in any topos $\MCE$, the subobject classifier $\top : 1 \rightarrowtail \Omega$ is an example of a dominance, and the classifying map $\Delta \to \Omega$ of any other dominance $\top_\Delta : 1 \rightarrowtail \Delta$ is monic, allowing a characterization of $\Delta$ as a special kind of subobject of $\Omega$.
\end{remark}

The terminology reflects the fact that a dominance can be used to give a class of \emph{domains} for partial maps.
Letting $\MCM_\Delta$ be the class of morphisms in $\MCE$ which arise as a pullback of $\top : 1 \rightarrowtail \Delta$, we see that $\MCM_\Delta$ consists entirely of monomorphisms, and all pullbacks of $\MCM_\Delta$-maps exist and are again in $\MCM_\Delta$.
Thus the construction of \Cref{def:m-partial-maps} applies to give a category $\pE_{\MCM_\Delta}$ of $\MCM_\Delta$-partial maps.

A particularly convenient setting is one where the pullback functor
$
  \top^* : \MCE / \Delta \to \MCE / 1 \simeq \MCE
$
between slice categories admits a right adjoint $\Pi_\top : \MCE \to \MCE / \Delta$.
In this case, writing $\Sigma_\Delta : \MCE / \Delta \to \MCE$ for the functor sending $f : A \to \Delta$ to $A$, we define $L_\Delta : \MCE \to \MCE$ as the composite $L_\Delta \coloneqq \Sigma_\Delta \circ \Pi_\top$.
\begin{lemma}[\cite{mulry-partial-map-classifiers-and-partial-cccs}, Thm.~2.4]
  In this setting, the functor $L_\Delta$ underlies a strong monad on $\MCE$ such that $\mathbf{Kl}(L_\Delta) \simeq \pE_{\MCM_\Delta}$.
\end{lemma}

In fact, the lifting monad $L_\Delta$ determines the dominance $\Delta$, since $\Delta \cong L_\Delta 1$ and $\top : 1 \rightarrowtail \Delta$ is the unit $\eta_1 : 1 \to L_\Delta1$.
In our applications, $\MCE$ has an initial object and the initial subobject $0 \rightarrowtail X$ is to be classifed by the dominance.

\begin{proposition}\label{prop:point-lifting}
  Let $\MCE$ be a category with an initial object $0$ and a dominance $\top : 1 \rightarrowtail \Delta$.
  The following are equivalent.
  \begin{enumerate}
  \item There is a map $\bot : 1 \rightarrowtail \Delta$ whose pullback with $\top$ is $0$.
  \item Every map $0 \to X$ is classified by $\Delta$.
  \item There exists a natural transformation $\bot : 1 \to L_\Delta$ from the constant functor with value $1$ to the lifting monad.
  \end{enumerate}
\end{proposition}


%% file: sheaves.tex
\subsection{Categories of sheaves on a site}
\label{sec:sheaves}

The notions of concrete site and concrete sheaf defined above in~\Cref{sec:concr-sites-sheav} are just special cases of the more general notions of \emph{site} and \emph{sheaf}.
The definitions and propositions in this subsection are standard (e.g.~\cite{John02}).

\begin{definition}
  For any small category $\bC$, the category of \emph{presheaves} is $\presh\bC \coloneqq [\bC^\op,\sets]$, the category of contravariant $\sets$-valued functors on $\bC$ and all natural transformations between them.
  The Yoneda embedding is denoted $y : \bC \to \presh\bC$.
\end{definition}

\begin{definition}
  A \emph{site} $(\bC,J)$ is a small category $\bC$ with a coverage $J$. A \emph{coverage} consists of, for every object $c\in\bC$, a set $J(c)$ of covering families $\{f_i:c_i\rightarrow c\}_{i\in I}$ satisfying the axiom (C) from \Cref{def:omnibus-concrete}.
\end{definition}

  A \emph{concrete site} (from~\Cref{def:omnibus-concrete}) is a site $(\bC,J)$ such that $\bC$ is a concrete category with terminal object $\star$ and initial object $0$, and $J$ satisfies ($\star$), (0), (M), and (L) (but see \Cref{rem:concrete-site-apology}). As in~\Cref{sec:concr-sites-sheav}, in a concrete site we define $\abs{c}=\bC(\star,c)$ and we can identify maps $f:c\rightarrow d$ with their action on points $\abs{f}:\abs{c}\rightarrow\abs{d}$.

Given a site $(\bC,J)$, a covering family $\{f_i:c_i\rightarrow c\}_{i\in I} \in J(c)$, and a presheaf $F\in\presh\bC$, a \emph{matching family} is a set $\{s_i\in F(c_i)\}_{i\in I}$ such that for all $i,j \in I$, $d \in \mathbb C$, $g : d \to c_i$, and $h : d \to c_j$ with $f_i\circ g=f_j\circ h$, we have $F(g)(s_i) = F(h)(s_j)$.

\begin{definition}
  Let $(\bC,J)$ be a site.
  A \emph{sheaf} on $(\bC,J)$ (or \emph{$J$-sheaf}) is a presheaf $F\in\presh\bC$ such that for every covering family $\{f_i:c_i\rightarrow c\}_{i\in I}$ and every matching family $\{s_i\in F(c_i)\}_{i\in I}$ there is a unique \emph{amalgamation} $s\in F(c)$ such that $F(f_i)(s)=s_i$ for all $i\in I$.
  The full subcategory of $\presh\bC$ whose objects are $J$-sheaves is denoted by $\sheaves(\bC,J)$.
\end{definition}

\begin{proposition}[e.g.~\cite{John02}, A4.1.8]
  The embedding $\sheaves(\bC,J)\to\presh\bC$ has a left adjoint $a : \presh\bC \to \sheaves(\bC,J)$ which preserves finite limits.
  $\sheaves(\bC,J)$ is a Grothendieck topos.
\end{proposition}

The left adjoint $a$ is called \emph{sheafification}.

\begin{remark}\label{rem:sheafified-representables-on-a-concrete-site}
  In general, the representable functors $y(c)$ for $c \in \bC$ are not sheaves (if the site is not subcanonical), so we instead use the sheafified representables $ay(c)$.
  To describe these, first note that, for any $c \in \bC$, the presheaf $\sets(|-|,|c|)$ is a sheaf and $y(c) \hookrightarrow \sets(|-|,|c|)$ is a subfunctor, where the component $y(c)(\star) \to \sets(|\star|,|c|)$ is a bijection.
  Since sheafification preserves monos, the sheafification of $y(c)$ is therefore given by closing the image of $y(c)(\star)$ under amalgations in $\sets(|-|,|c|)$.
  Using the (M) and (L) axioms, we can write
  \begin{multline*}
    ay(c)(d) \cong \big\{ \phi \in \sets(|d|,|c|) \midd \exists \{f_i : d_i \to d\} \in J(d).\\ \forall i. \phi \circ |f_i| \in \im\big(y(c)(d_i) \hookrightarrow \sets(|d_i|,|c|)\big) \big\},
  \end{multline*}
  i.e.\ $ay(c)(d)$ is isomorphic to the set of functions $|d| \to |c|$ which are \emph{$J$-locally} given by morphisms into $c$ in $\bC$.
\end{remark}

\begin{definition}
  Let $(\bC,J)$ be a concrete site. A \emph{concrete presheaf} is a presheaf $F : \presh\bC$ such that, for every $c \in \mathbb C$, the function $\langle F(x : \star \to c)\rangle_{x \in |c|} : F(c) \to \Set(\abs{c}, F(\star))$ is injective. 
\end{definition}

\begin{proposition}[e.g.~\cite{baez-hoffnung-smooth,dubuc-concrete-quasitopoi}]
  Let $(\bC,J)$ be a concrete site and $F$ a concrete presheaf which is also a sheaf. The functor that sends $F$ to the concrete sheaf $X$ (in the sense of~\Cref{def:concr-sheav}) given by the set $\abs{X}=F(\star)$ with ${R^c_X=\mathsf{Im}(\langle F(x : \star \to c)\rangle_{x \in |c|})}$ is an equivalence.
\end{proposition}

\begin{remark}
  The argument in \Cref{rem:sheafified-representables-on-a-concrete-site} shows that representable functors on a concrete site are concrete presheaves, and that the sheafified representables are still concrete.
  By a similar argument, the sheafification functor sends any concrete presheaf to a concrete sheaf.
\end{remark}

For a presheaf $X\in\presh\bC$ write $\abs{X}$ for $\bC(\star,X)$. We can think of a concrete presheaf $X$ as being the set $\abs{X}$ together with a set of functions $\abs{c}\rightarrow\abs{X}$ for each $c\in\bC$. A natural transformation $\alpha : Y\rightarrow X$ from a presheaf $Y$ to a concrete presheaf $X$ is determined by the function $\alpha_\star:\abs{Y}\rightarrow\abs{X}$.

\begin{remark}
  The category of concrete sheaves $\conc(\bC,J)$ forms a (Grothendieck) \emph{quasitopos}.
  It is still cartesian closed.
\end{remark}

\begin{proposition}[e.g.~\cite{John02}, \S C2.2; \cite{baez-hoffnung-smooth}]\label{prop:conc-exp-ideal}
  Let $(\bC,J)$ be a concrete site.
  The full inclusion $\conc(\bC,J) \to \sheaves(\bC,J)$ preserves all limits, exponentials, and coproducts, and has a left adjoint. 
\end{proposition}

\subsection{Dominances and pre-admissible monos on a site}
\label{sec:preadmissible}

Let $(\bC,J)$ be a site, not necessarily concrete.
There is a more general version of~\Cref{def:class-of-admissible-monos}, as follows.
Suppose $\MCM$ is a class of monomorphisms in $\bC$ satisfying the following.
\begin{enumerate}
\item $\MCM$ contains all the isomorphisms and is closed under composition.
\item All pullbacks of $\MCM$-maps exist and are again in $\MCM$.
\end{enumerate}
Then there is a presheaf $\Delta_\MCM \in\presh\bC$ given by $\Delta_\MCM(c) \coloneqq \sub_\MCM(c)$, the set of isomorphism classes of $\MCM$-subobjects, with functorial action $\Delta_\MCM(f : a \to c) : \Delta_\MCM(c) \to \Delta_\MCM(a)$ given by pullback.
\begin{definition}
  $\MCM$ is a class of \emph{pre-admissible} monomorphisms in $\bC$  (for $(\bC,J)$) if it satisfies the two conditions above and the $\Delta_\MCM$ is a $J$-sheaf.
\end{definition}

There is a map
$
   \top : 1 \rightarrowtail \Delta_\MCM
$
given by $\top_c(\star) = [1_c] \in \sub_\MCM(c)$.
The following generalizes Theorem 2.6 of \cite{mulry-partial-map-classifiers-and-partial-cccs}, which covers only the case where $J$ is a trivial coverage.

\begin{theorem}
  Let $(\bC,J)$ be a site with $\MCM$ a  class of pre-admissible monos.
  Then the map $\top : 1 \to \Delta_\MCM$ is a dominance in $\sheaves(\bC,J)$, and $\sheaves(\bC,J)(ay(c),\Delta_\MCM) \cong \sub_\MCM(c)$.
\end{theorem}
\begin{proof}
  Since $\Delta_\MCM$ is a sheaf we have $\sheaves(\bC,J)(ay(c),\Delta_M) \cong [\bC^\op,\sets](y(c),\Delta_M) \cong \Delta_M(c) \cong \sub_\MCM(c)$.
  Moreover, if $\chi : ay(c) \to \Delta_\MCM$ corresponds to an $\MCM$-subobject $m : c' \rightarrowtail c$, then the pullback of $\top : 1 \rightarrowtail \Delta_\MCM$ along $\chi$ is $ay(m) : ay(c') \to ay(c)$, since the sheafification $a$ preserves finite limits.

  It is easy to see that a subobject $m : X' \rightarrowtail X$ arises as a pullback of $\top : 1 \rightarrowtail \Delta_\MCM$ along some map $X \to \Delta_\MCM$ iff $m$ is `representably in $\MCM$', meaning that the pullback of $m$ along any map $ay(c) \to X$ from a (sheafified) representable has the form $ay(m') : ay(c') \to ay(c)$ for some $m' \in \sub_\MCM(c)$.
  From this description it follows easily that the subobjects classified by $\Delta_\MCM$ are closed under composition, as required.
\end{proof}

We can spell out the formula from~\Cref{sec:dominance} for the lifting monad in this case for $X \in \sheaves(\bC,J)$:
\begin{displaymath}
  L_\MCM(X)(c) = \coprod_{(m : c' \rightarrowtail c) \in \sub_\MCM(c)} X(c')
\end{displaymath}
where the sum is over isomorphism classes of $\MCM$-subobjects of $c$.

\subsubsection{Concreteness}\label{sec:concreteness}

  Let $\MCM$ be a class of pre-admissible monos in a concrete site $(\bC,J)$. We are interested in the case where the dominance $\Delta_\MCM$ is a \emph{concrete sheaf}.

This means each $\MCM$-subobject $(m : c' \rightarrowtail c)$ of each $c \in \bC$ is determined by the subset $|m| \subseteq |c|$ of points of $c$ that factorize through them.
It is straightforward to see that the order-relation $m \leq m'$ between $\MCM$-subobjects is now also reflected by the relation $|m| \subseteq |m'|$. Notice that in this case, $\Delta_\MCM$ looks like the  class of admissible monos from~\Cref{sec:admiss-monos-lift} but without the $\omega$-cpo structure.

Assume further that every map in $0\rightarrow c$ is in $\MCM$.
In this case, it can be shown that because $\Delta_\MCM$ is concrete, $\Delta_\MCM(\star)$ has exactly two elements, $[1_\star]$ and $[0\rightarrow\star]$.
Therefore, the lifting of a sheaf $X\in\sheaves(\bC,J)$, $L_\MCM X$, has the set of points:
\begin{displaymath}
  |L_\MCM(X)| \cong \coprod_{(c \rightarrowtail \star) \in \sub_\MCM(\star)} X(c) \cong X(\star) + X(0) \cong |X| + 1
\end{displaymath}
Since $X(0)\cong 1$ because $X$ is a sheaf. Notice that this is the same underlying set of points as that of the lifting monad from~\Cref{sec:admiss-monos-lift}.

\begin{lemma}\label{lem:lift-preserv-conc}
  Let $(\bC,J)$ be a concrete site with a class $\MCM$ of pre-admissible monos such that all maps $0\rightarrow c$ are in $\MCM$, and $\Delta_\MCM$ is concrete. Then the lifting monad $L_\MCM$ preserves concreteness.
\end{lemma}
This result will be used in the proof of Theorem~\ref{thm:sheaves-adequate}.

\begin{proposition}
  Let $(\bC,J)$ be a concrete site with a  class $\MCM$ of pre-admissible monos, such that all maps $0\rightarrow c$ are in $\MCM$.
  Then the dominance $\Delta_\MCM$ classifies the subobject $0\rightarrow 1$ from $\sheaves(\bC,J)$.
\end{proposition}
It then follows from~\Cref{prop:point-lifting} that the monad $L_\MCM$ has a point i.e.~a natural transformation $\bot:1\rightarrow L_\MCM$. This fact will be used in the next section.


%% file: recursion.tex
\section{Recursion in categories of sheaves}
\label{sec:recursion}
We now recall how recursion can be understood once partiality is set up (as in \S\ref{sec:beginning-of-sdt}). We do this by specializing some general ideas from synthetic domain theory to sheaf categories, following e.g.~\cite{matache-moss-staton-fscd-2021}. 

Let us consider any small site $(\bC,J)$ with a terminal object $\star$ and initial object $0$ covered by the empty family, with a class of pre-admissible monomorphisms $\MCM$ such that every map $! : 0 \to c$ is in $\MCM(c)$.
It is interesting to note that for each sheaf $X \in \sheaves(\bC,J)$ the points of $X$ carry an intrinsic \emph{information ordering}, given by the image of
\begin{displaymath}
  \nattrans(\Delta_\MCM,X) \to \sets(|\Delta_\MCM|,|X|) \to \sets(\{\bot,\top\},|X|) \to |X|^2,
\end{displaymath}
where we are using the fact that the assumptions on $(\bC,J)$ and $\MCM$ provide that $\Delta_\MCM$ has at least two points, classifying the bottom and top elements of $\sub(\star)$.
This relation is necessarily reflexive, but there is no reason for it to be transitive or antisymmetric in general.




It is common in denotational semantics for a recursively defined term to denote a limit or supremum of an ascending sequence of approximations.
In the absence of transitivity, it makes sense to consider intrinsic notions of `increasing sequence' and `limit of an increasing sequence'.
The approach given here is essentially a variation on that of~\cite{fiore-plotkin-an-extension-of-models-of-adt-to-models-of-sdt}.

Let $\omega_\MCM = \omega$ be the colimit in $\sheaves(\bC,J)$ of the diagram
\begin{equation}\label{eq:1}
  1 \xrightarrow{\bot_1} L1 \xrightarrow{L(\bot_1)} LL1 \xrightarrow{LL(\bot_1)} \ldots
\end{equation}
and $\wbar_\MCM = \wbar$ the limit in $\sheaves(\bC,J)$ of the diagram
\begin{equation}\label{eq:2}
  1 \xleftarrow{!} L1 \xleftarrow{L!} LL1 \xleftarrow{LL!} \ldots.
\end{equation}
There is an evident comparison map $i : \omega \to \wbar$.


\begin{lemma}
  $|i| : |\omega| \to |\wbar|$ is given by $\bN \hookrightarrow \bN \cup \{\infty\}$, and all maps $\omega \to X$ or $\wbar \to X$ are monotone from the natural order on $\bN \cup \{\infty\}$ to the intrinsic information order on $X$.
\end{lemma}

It is also straightforward to construct maps $\succe_\omega : \omega \to \omega$, $\succe_\wbar : \wbar \to \wbar$ and $\infty : 1 \to \wbar$
with the obvious action on points such that $  (i \circ \succe_\omega = \succe_\wbar \circ i)$ and $(\succe_\wbar \circ \infty = \infty)$.

\subsection{Completeness and fixed points}

In synthetic domain theory, one restricts to a subcategory of `complete' objects where fixed point operators can be defined. If we think of a morphism $\omega\rightarrow X$ as a chain in $X$, completeness implies that such a chain has a least upper bound.

Recall that an object $X$ is said to be \emph{right-orthogonal} to a morphism $f:A\rightarrow B$ if every map $A\rightarrow X$ factors uniquely through $f$.
In this situation we write $f\perp X$.

Denote by $\omega^\mathsf{P}$ the colimit of diagram~(\ref{eq:1}) in $\presh{\bC}$.
Notice that in general $\omega^\mathsf{P}$ is not a $J$-sheaf.
The limit of diagram~(\ref{eq:2}) in $\presh{\bC}$ is the same as in sheaves, i.e.~$\wbar$.
As before, let $i^\mathsf{P}: \omega^\mathsf{P} \rightarrow \wbar$ be the evident comparison map.
The equivalence between~\ref{item:1} and~\ref{item:4} below is quite standard, see e.g.~\cite{fiore-plotkin-an-extension-of-models-of-adt-to-models-of-sdt}.

\begin{lemma}\label{lem:internally-orthogonal}
  Let $X \in \sheaves(\bC,J)$.
  The following are equivalent.
  \begin{enumerate}
  \item\label{item:1} The map $X^i : X^\wbar \to X^\omega$ is an isomorphism.
  \item\label{item:4} For all $A \in \sheaves(\bC,J)$, $(i \times 1_A : \omega \times A \to \wbar \times A) \perp X$.
  \item For all $c \in \bC$, $(i^\mathsf{P} \times 1_{y(c)} : \omega^\mathsf{P} \times y(c) \to \wbar \times y(c)) \perp X$.
  \end{enumerate}
\end{lemma}

\begin{definition}\label{def:complete}
  Consider a site $(\bC,J)$ with a class of pre-admissible monos $\MCM$. A sheaf $X \in \sheaves(\bC,J)$ is:
  \begin{itemize}
  \item \emph{$L_\MCM$-complete} if $X$ satisfies the conditions of \Cref{lem:internally-orthogonal}, 
  \item \emph{well-complete} if $L_\MCM X$ is $L_\MCM$-complete.
  \end{itemize}
\end{definition}

The present abstract setting admits the following fixed point theorem. The theorem is about well-complete objects with respect to $L_\MCM$, that are moreover $L_\MCM$-algebras (i.e.~objects $X$ equipped with a morphism $L_\MCM(X)\to X$ satisfying conditions).

\begin{theorem}[\cite{matache-moss-staton-fscd-2021}]\label{thm:fixed-points-x}
  Let $X \in \sheaves(\bC,J)$ be a well-complete object that is also an $L_\MCM$-algebra . Then for any map $g : \Gamma \times X \rightarrow X$ we can construct a fixed point $\phi_g: \Gamma \rightarrow X$ such that $\phi_g(\rho)=g(\rho,\phi_g(\rho))$.
\end{theorem}

\begin{corollary}[\cite{matache-moss-staton-fscd-2021}]\label{cor:cbv-fixed-points}
  Consider objects $\Gamma$, $A$, $B$ in $\sheaves(\bC,J)$ such that $(L_\MCM B)^A$ is a well-complete object. Then there is a fixed point combinator $\big((L_\MCM B)^A\Rightarrow (L_\MCM B)^A\big)\rightarrow (L_\MCM B)^A$.
\end{corollary}

We will use~\Cref{cor:cbv-fixed-points} to interpret fixed points suitable for call-by-value.

\subsection{The subcategory of well-complete objects}\label{sec:subc-compl-objects}

We now explore conditions on $\Delta_\MCM$ that guarantee a supply of $L_\MCM$-complete objects sufficient to model \pcfv\ using~\Cref{cor:cbv-fixed-points}.
Later in~\Cref{sec:comb-sites-admiss} we will translate conditions to the site $(\bC,J)$ and class $\MCM$ of pre-admissible monos.

As in~\cite{matache-moss-staton-fscd-2021}, we consider a slight strengthening of the $L_\MCM$-completeness condition, which roughly says that an object is $L_\MCM$-complete with respect to partial maps.
\begin{definition}
  Let $\om$ be the class of maps in $\sheaves(\bC,J)$ which are pullbacks of maps $i \times 1_A : \omega_\MCM \times A \to \wbar_\MCM \times A$ along subobjects of $\wbar_\MCM \times A$ classified by $\Delta_\MCM$.
  Write $\om^\boxslash$ for the class of objects right orthogonal to every map in $\om$.
\end{definition}
The following facts and proposition are explained in~\cite{matache-moss-staton-fscd-2021}:
\begin{itemize}
\item $\om$ is closed under the operations $(-)\times 1_A$, and under pullback along subobjects classified by $\Delta_\MCM$.
\item $\om^\boxslash$ is contained in the class of $L_\MCM$-complete objects.
\item $\om^\boxslash$ is a reflective subcategory of $\sheaves(\bC,J)$, is closed under limits in $\sheaves(\bC,J)$, and is an exponential ideal.
\end{itemize}

\begin{proposition}\label{prop:odeltaclass-omnibus-proposition}
  Suppose that $\Delta_\MCM$ is $L_\MCM$-complete.
  \begin{itemize}
  \item $\Delta_\MCM$ is in $\om^\boxslash$, and for $A \in \sheaves(\bC,J)$, $A \in \om^\boxslash$ iff $A$ is well-complete iff $L_\MCM A \in \om^\boxslash$.
  \item $\om^\boxslash$ is closed under $L_\MCM$ and contains $0$.
  \item $\om^\boxslash$ is closed under $I$-indexed coproducts iff $\sum_{I'} 1 \in \om^\boxslash$ for some set $I'$ with $\abs{I} \leq \abs{I'}$.
  \end{itemize}
\end{proposition}
We will use Prop.~\ref{prop:odeltaclass-omnibus-proposition} in \S\ref{sec:sheaf-model-pcfv}, to give an interpretation of \pcfv.
In Propositon~\ref{prop:lnat-complete}, we explore properties of $\MCM$ that will allow us to deduce the $L_\MCM$-completeness of $\Delta_\MCM$.

\subsection{The vertical natural numbers}
\label{sec:vset}
Here we recall the `canonical' example of a site with pre-admissible monos such that $\Delta_\MCM$ is $L_\MCM$-complete.
This is essentially the same as the category $\MCH$ considered in \cite{fiore-rosolini-h} as a model of synthetic domain theory, except we omit their coverage which plays no role for us.

\begin{definition}
  Let $\vseq \in \wcpo$ be the ordinal $\omega + 1 \cong \bN \cup \{\infty\}$ considered as an $\wcpo$.
  Let $\vertendo$ be the full subcategory of $\wcpo$ with just the object $\vseq$.
  We define $\vsets \coloneqq \presh\vertendo$.
\end{definition}

Although this is the most convenient description of $\vsets$ as a plain category, it is necessary to extend the site a little in order to get the correct class of pre-admissible monos.
Let $\vertendo_0$ be the full subcategory of $\wcpo$ whose objects are $\vseq$, the terminal object $\star$, and the intial object $0$.
Let $J_\vseq$ be the coverage with
\begin{mathpar}
  J_\vseq(\vseq) = \{\{1_\vseq\}\}
  \and
  J_\vseq(\star) = \{\{1_\star\}\}
  \and
  J_\vseq(0) = \{\emptyset,\{1_0\}\}.
\end{mathpar}
Then it is easy to see that $\sheaves(\vertendo_0,J_\vseq) \simeq \vsets$.
Explicitly, the equivalence sends $X \in \presh\vertendo$ to the sheaf $\overline X$ where
\begin{mathpar}
  \overline X(\vseq) = X(\vseq)
  \and
  \overline X(0) \cong 1
  \and
  \overline X(\star) = \presh{\vertendo}(1,X) 
\end{mathpar}
and the obvious functorial action.
Now consider the following class $\MCM_\vseq$ of monomorphisms in $\vertendo_0$.
\begin{mathpar}
  \MCM_\vseq(\vseq) = \{ (\lambda x.x + n) \in \vertendo_0(\vseq,\vseq) \mid n \in \bN \} \cup \{ ! : 0 \to \vseq \}
  \and
  \MCM_\vseq(0) = \{ ! : 0 \to 0 \}
  \and
  \MCM_\vseq(\star) = \{ 1_\star : 1 \to 1, ! : 0 \to 1 \}
\end{mathpar}

\begin{lemma}
  $(\vertendo_0,J_\vseq)$ is a concrete site, and $\MCM_\vseq$ is a class of pre-admissible monos.
\end{lemma}

Writing $\Delta_\vseq$ for $\Delta_{\MCM_\vseq}$ and $L_\vseq$ for $L_{\MCM_\vseq}$, our main interest in $\vsets$ is the following, which allows us to apply Prop.~\ref{prop:odeltaclass-omnibus-proposition}.

\begin{proposition}[\cite{matache-moss-staton-fscd-2021}, Lemma~5.3]
  $\Delta_\vseq$ is $L_\vseq$-complete.
\end{proposition}


%% file: sheaf-model.tex
\section{Sheaf models of PCF with adequacy}\label{sec:sheaf-model-pcfv}

In this final section we explain when a concrete site $(\bC,J)$ together with a class of admissible monos (Def.~\ref{def:class-of-admissible-monos}) gives an adequate sheaf model of \pcfv\  (Thm.~\ref{thm:sheaves-adequate}). We do this by combining the site $\bC$ with the site of vertical natural numbers from (\S\ref{sec:vset}). We also connect this sheaf-based model back to the $\omega$-concrete sheaves of Section~\ref{sec:modelling-recursion} (Prop.~\ref{prop:wconc-sheav-embed}).

\subsection{Combining sites and admissible monos}\label{sec:comb-sites-admiss}
\begin{proposition}
  Consider a concrete site $(\bC,J)$.
  A class of pre-admissible monos $\MCM$ is a class of \emph{admissible} monos (in the sense of~\Cref{def:class-of-admissible-monos}) if:
  \begin{enumerate}
  \item Every map $0\rightarrow c$ is in $\MCM$.
  \item $\Delta_\MCM$ is \emph{concrete}. We saw in~\Cref{sec:concreteness} that this means $\MCM$-subobjects $(m:c'\rightarrowtail c)$ are determined by the set of points of $c$ that factorize through them, $\abs{m}\subseteq\abs{c}$, and the order $m\leq m'$ is given by inclusion $\abs{m}\subseteq\abs{m'}$.
  \item\label{item:2} For every increasing chain of monos on $c$, $(m_n:c_n\rightarrowtail c)_{n\in\bN} \in\MCM$, the subobject $m_\infty:c_\infty\rightarrowtail c$ determined by the set of points $\bigcup_{n\in \bN}\abs{m_n}$ is in $\MCM$.
  \item\label{item:3} Given an increasing chain of monos $(m_n:c_n\rightarrowtail c)_{n\in\bN} \in\MCM$, the closure under precomposition (with any morphism) of the set $\{m_n:c_n\rightarrowtail c_\infty\}_{n\in\bN}$ contains a covering family of $c_\infty$.
  \end{enumerate}
\end{proposition}

\begin{example}
  The  class of pre-admissible monos $\MCM_\vseq$ from $(\vertendo_0,J_\vseq)$ is a class of admissible monos.
\end{example}

\begin{lemma}\label{lem:comb-sites}
  Let $(\bC_1,J_1,\MCM_1)$ and $(\bC_2,J_2,\MCM_2)$ be two concrete sites with classes of admissible monos. Let $\bC_1+\bC_2$ be the category obtained from $\bC_1$ and $\bC_2$ by identifying the respective terminal objects and the respective initial objects, and adding all constant maps between all objects. Then $(\bC_1+\bC_2, J_1\cup J_2, \MCM_1\cup\MCM_2)$ is also a concrete site with a class of admissible monos. 
\end{lemma}

In order to model recursion, we want to find a sheaf category where $\Delta_\MCM$ and $L_\MCM(\sum_0^\infty1)$ are $L_\MCM$-complete objects. The next proposition shows that using the site $\vertendo_0$ and a class of admissible monos we can obtain such a sheaf category:

\begin{proposition}\label{prop:lnat-complete}
Let $(\bC,J,\MCM)$ be a concrete site with a class of admissible monos. In the sheaf category $\sheaves(\bC+\vertendo_0,J\cup J_\vseq)$ the dominance $\Delta_{\MCM\cup\MCM_\vseq}$ and $L_{\MCM\cup\MCM_\vseq}(\sum_0^\infty1)$ are $L_{\MCM\cup\MCM_\vseq}$-complete objects. 
\end{proposition}
\begin{proof}[Proof sketch]
  First show that $\Delta_{\MCM\cup\MCM_\vseq}$ is right-orthogonal to $i^\mathsf{P}:\omega^\mathsf{P}\rightarrow\wbar$.
  From~\Cref{lem:lift-preserv-conc}, the lifting monad $L_{\MCM\cup\MCM_\vseq}$ preserves concreteness so $L_{\MCM\cup\MCM_\vseq}(1)(\star)\cong\Delta_{\MCM\cup\MCM_\vseq}(\star)\cong\{0\leq 1\}$.
  Using the fact that $\vseq$ is part of the site, and the colimit description of $\omega^\mathsf{P}$, we can show that the maps $f:\omega^\mathsf{P}\rightarrow \Delta_{\MCM\cup\MCM_\vseq}$ are  the infinite monotone binary sequences.
  This gives a candidate extension of $f$ to $\overline{f}:\wbar\rightarrow \Delta_{\MCM\cup\MCM_\vseq}$ which we show is natural and unique. For uniqueness use the fact that $\wbar$ is a limit and that $\vseq$ is part of the site.

  From~\Cref{lem:internally-orthogonal}, to show $\Delta_{\MCM\cup\MCM_\vseq}$ is $L_{\MCM\cup\MCM_\vseq}$-complete it is enough to show that every map $f:\omega^\mathsf{P}\times yc \rightarrow \Delta_{\MCM\cup\MCM_\vseq}$ can be extended to $\wbar\times yc$ for any object $c$ in $\bC+\vertendo_0$.
  Using the Yoneda lemma we can describe maps $f:\omega^\mathsf{P}\times yc \rightarrow \Delta_{\MCM\cup\MCM_\vseq}$ as increasing chains of $(\MCM\cup\MCM_\vseq)$-subobjects of $c$.
  Condition~(\ref{item:2}) in the definition of class of admissible monos ensures there is a sup for the chain $f$, which we can show defines a natural extension $\overline{f}$.
  For uniqueness we use the fact that each $f(-,x):\omega^\mathsf{P}\rightarrow \Delta_{\MCM\cup\MCM_\vseq}$ has a unique extension.

  Following the same pattern, we first show $f:\omega^\mathsf{P}\rightarrow L_{\MCM\cup\MCM_\vseq}(\sum_0^\infty1)$ has a unique extension, using  the fact that each map $\omega^\mathsf{P}\rightarrow L_{\MCM\cup\MCM_\vseq}(\sum_0^\infty1)$ or $\wbar\rightarrow L_{\MCM\cup\MCM_\vseq}(\sum_0^\infty1)$ factors through some $L_{\MCM\cup\MCM_\vseq}(1)\rightarrowtail L_{\MCM\cup\MCM_\vseq}(\sum_0^\infty1)$.

  Next, notice that maps $f:\omega^\mathsf{P}\times yc\rightarrow L_{\MCM\cup\MCM_\vseq}(\sum_0^\infty1)$ can be described as an increasing chain of $(\MCM\cup\MCM_\vseq)$-subobjects of $c$, $(c_n\rightarrowtail c)_{n\in\bN}$, together with a chain of functions $(g_n:\abs{c_n}\rightarrow\bN)_{n\in\bN}$ (each extending the previous one), such that for each $n$ there is a cover of $c_n$ on which $g_n$ is locally constant.
  Condition~(\ref{item:3}) in the definition of class of admissible monos and axiom (L) of $J\cup J_\vseq$ guarantee that $g_\infty=\bigcup_{n\in\bN}g_n:\abs{c_\infty}\rightarrow\bN$ is locally constant on a cover of $c_\infty$.

  Thus we have a candidate extension $\overline{f}$ of type $\wbar\times yc\rightarrow L_{\MCM\cup\MCM_\vseq}(\sum_0^\infty1)$.
  To show naturality we prove that $\overline{f}$ factors through the map $L_{\MCM\cup\MCM_\vseq}(1) \times yc \rightarrow L_{\MCM\cup\MCM_\vseq}(\sum_0^\infty1)$ given at $\star$ by the function $\phi:\{0\leq 1\}\times \abs{c}\rightarrow \bN+\{\bot \}$,
  $\phi(0,x)=\bot$, $\phi(1,x)= g_\infty(x)$ if $x\in\abs{c_\infty}$ or $\bot$ otherwise,       
which we can show is natural directly.
\end{proof}

The model of \pcfv\ from~\Cref{sec:denot-semant-pcfv} is closely related to the category of sheaves $\sheaves(\bC+\vertendo_0,J\cup J_\vseq)$:

\begin{proposition}\label{prop:wconc-sheav-embed}
Let $(\bC,J,\MCM)$ be a concrete site with a class of admissible monos. There is a functor $F:\omega\conc(\bC,J)\rightarrow \sheaves(\bC+\vertendo_0,J\cup J_\vseq)$ which is full, faithful, preserves products, coproducts and exponentials, and commutes with the lifting monad i.e.~$FL_\MCM = L_{\MCM\cup\MCM_\vseq}F$. Moreover, for every $\omega$-concrete sheaf $X$, $FX$ is a concrete $L_{\MCM\cup\MCM_\vseq}$-complete sheaf.  
\end{proposition}
\begin{proof}[Proof notes]
  The interesting part in the definition of $F$ is:
    $(FX)(\vseq) = \big\{ f:\abs{\vseq}\rightarrow\abs{X} \midd f\text{ an }\omega\text{-chain with sup, in }\abs{X} \big\}.$
  Otherwise, $F$ leaves $X$ unchanged.
\end{proof}
Given the embedding from Prop.~\ref{prop:wconc-sheav-embed} we can deduce the fixed point construction from Prop.~\ref{prop:wconc-fp}, using~\Cref{cor:cbv-fixed-points}.

\begin{example}[Fully abstract model of \pcfv\ \cite{matache-moss-staton-fscd-2021}]\label{ex:fscd}
  The fully abstract model of \pcfv\ from~\cite{matache-moss-staton-fscd-2021} is presented by a concrete site with a class of admissible monos $(\bC,J,\MCM)$.

  Roughly speaking, to construct $(\bC,J,\MCM)$ start from a concrete site and a class of admissible monos $(\ssp,J_\ssp,\MCM_\ssp)$ whose definition we omit. Intuitively \ssp\ is chosen to encode \pcfv-definable functions between ground types.
  As explained in~\Cref{def:m-partial-maps}, there is a category of partial maps, $\ssp_\bot$, with domains in $\MCM_\ssp$. For every faithful functor $F:\MCC\rightarrow\ssp_\bot$, we can construct another concrete site $(\MCI_{\MCC,F},J_{\MCC,F},\MCM_{\MCC,F})$ where the objects are pairs $(c\in\MCC,\ U\rightarrowtail Fc \in\MCM_\ssp)$; a (total) morphism $(c,\ U\rightarrowtail Fc)\rightarrow(c',\ U'\rightarrowtail Fc')$ is either constant or comes from a partial map $F\phi$ with domain $U$. Thus $\MCI_{\MCC,F}$ is a ``totalization'' of $F:\MCC\rightarrow\ssp_\bot$, where each partial map is represented by a total one. The coverage $J_{\MCC,F}$ and class of monos $\MCM_{\MCC,F}$ are obtained by restricting $J_\ssp$ and $\MCM_\ssp$ appropriately.

By combining the $(\MCI_{\MCC,F},J_{\MCC,F},\MCM_{\MCC,F})$ sites for all $F:\MCC\rightarrow\ssp_\bot$ using~\Cref{lem:comb-sites} we obtain $(\bC,J,\MCM)$. Then the sheaf category $\sheaves(\bC+\vertendo_0,J\cup J_\vseq)$ is exactly the model $\MCG$ from~\cite{matache-moss-staton-fscd-2021}, and $\MCM\cup\MCM_\vseq$ induces the same lifting monad
.
  
\end{example}

\subsection{Adequacy}
Given the concrete site $(\bC,J,\MCM)$, we interpret \pcfv\ in the sheaf category $\sheaves(\bC+\vertendo_0,J\cup J_\vseq)$ using the lifting monad $L_{\MCM\cup\MCM_\vseq}$ obtained from the class of admissible monos $\MCM\cup\MCM_\vseq$.
The type $\nat$ is interpreted using the infinitary coproduct $\sum_0^\infty1$; the other type constants $\alpha$ are interpreted by concrete sheaves $\den\alpha$.
The rest of the interpretation is defined using the structure of the category, similarly to~\Cref{sec:denot-semant-pcfv}.

Assuming that the type constants $\den\alpha$ are well-complete, \Cref{prop:lnat-complete} and~\Cref{prop:odeltaclass-omnibus-proposition}, and its preceding discussion imply that all \pcfv\ types are $L_\MCM$-complete objects.
Hence,  we can use the construction of fixed points from~\Cref{cor:cbv-fixed-points} to interpret $(\rec{f}{x}{t})$. We are now able to state and prove the main theorem of the paper: 

\begin{theorem}[Adequacy]\label{thm:sheaves-adequate}
  A concrete site with a class of admissible monos, $(\bC,J,\MCM)$, presents a sound and adequate model, in $\sheaves(\bC+\vertendo_0,J\cup J_\vseq)$, and in $\conc(\bC+\vertendo_0,J\cup J_\vseq)$, of \pcfv.
\end{theorem}
\begin{proof}[Proof sketch]
  The ground types $\nat$ and $\alpha$ are interpreted as \emph{concrete} sheaves, $L_{\MCM\cup\MCM_\vseq}$ preserves concreteness (\Cref{lem:lift-preserv-conc}), and the concrete sheaves are an exponential ideal (\Cref{prop:conc-exp-ideal}). So all types are concrete sheaves. Therefore, morphisms between them are determined by the underlying function at $\star$. This means that both soundness and adequacy can be proved following the same strategy as in the cpo model of \pcfv\ (e.g.~\cite[Lemma~11.14]{winskel-semantics}). Soundness is proved by induction on the definition of $\Downarrow$.
  
  For adequacy, we define a logical relation using the set of points of each value and computation: $\logval{\tau} \subseteq \abs{\den\tau} \times \mathsf{Val}_\tau$ and $\logcomp{\tau} \subseteq \abs{L_\MCM\den\tau} \times \mathsf{Comp}_\tau$. (Where $\mathsf{Val}_\tau$ is the set of values of type $\tau$, and similarly for computations.)
  \begin{align*}
    \logval{\tau\rightarrow\tau'} &= \big\{ (d,v) \midd \forall a {\in} \abs{{\den\tau}},\,w \in \mathsf{Val}_\tau.\, a \logval{\tau} w \Rightarrow (d\ a) \logcomp{\tau'} (v\ w)\big\} \\
    \logcomp{\tau} &= \big\{ (d,t) \midd  \big(d=\abs{\eta_{\den\tau}}\circ d'\big) \implies \exists w.\ t\Downarrow w ,\ d'\logval{\tau} w \big\}
  \end{align*}
  and for a type constant $\logval{\alpha}$ the identity relation. The relation specifies when a term is approximated by an element of the model.

  The `fundamental property' is proved by induction on terms.
  For the $\mathsf{rec}$ case we prove by induction on types that all subobjects of the form $\{(-) \logcomp{\tau''} t''\}$ are closed under sups of chains.
  (Here a chain is a map $\omega\rightarrow L_{\MCM\cup\MCM_\vseq}\den{\tau''}$, and a chain with a lub is $\wbar\rightarrow L_{\MCM\cup\MCM_\vseq} \den{\tau''}$.) This replaces the proof from cpo's that the logical relation is an admissible subset. The $\mathsf{let}$ case works the same as in cpo because the lifting monad acts on the underlying sets in the same way (see~\Cref{sec:modelling-recursion}).
\end{proof}

The adequacy proof above extends easily when we add product and sum types as in~\Cref{app:typing:rules}.

Given the embedding from~\Cref{prop:wconc-sheav-embed} we can finally deduce the adequacy result for $\omega\conc(\bC,J)$ from~\Cref{thm:wconc-adequate}, using~\Cref{thm:sheaves-adequate}.


%% file: app-typing.tex
\section{\pcfv\ extended with products and sums}\label{app:typing:rules}

\begin{figure*}[b]\framebox{\begin{minipage}{\linewidth}\emph{Typing rules:}
   \[\begin{gathered}
  \inferrule{ }{\Gamma \vterm \star : \tone} \quad
  \inferrule{ \Gamma \vterm v :\tau}{\Gamma \vterm \inl{v} :\tau +\tau'} \quad
  \inferrule{ \Gamma \vterm v :\tau'}{\Gamma \vterm \inr{v} :\tau +\tau'} \\
  \inferrule{ }{\Gamma,x:\tau,\Gamma' \vterm x :\tau} \quad
  \inferrule{ }{\Gamma \vterm \vzero :\nat} \quad
  \inferrule{\Gamma\vterm v: \nat}{\Gamma \vterm \suc{v}: \nat}\\
  \inferrule{\Gamma,\,x:\tau \cterm t:\tau'}{\Gamma \vterm \lbd{x}{t} :\tau\rightarrow\tau'} \quad
  \inferrule{\Gamma,\, f:\tau\rightarrow\tau',\, x:\tau \cterm t:\tau'}{\Gamma \vterm \rec{f}{x}{t} : \tau\rightarrow \tau'} \\
  \inferrule{\Gamma \vterm v:\tau \\ \Gamma \vterm v':\tau'}{\Gamma \vterm (v,v'):\tau\times\tau'} \quad  \inferrule{\Gamma \vterm v: \tau\times\tau'}{\Gamma \cterm \pi_1v : \tau} \quad
  \inferrule{\Gamma \vterm v: \tau\times\tau'}{\Gamma \cterm \pi_2v : \tau'} 
\end{gathered}\qquad\begin{gathered}
  \inferrule{\Gamma \vterm v :\tau+\tau' \\ \Gamma,x:\tau \cterm t : \sigma \\ \Gamma,y:\tau' \cterm t':\sigma}{\Gamma \cterm \casesum{v}{x}{t}{y}{t'} :\sigma} \\
  \inferrule{\Gamma \vterm v : \tzero}{\Gamma \cterm \caseempty v : \tau} \quad
  \inferrule{\Gamma \vterm v :\tau\rightarrow \tau' \\ \Gamma \vterm w :\tau}{\Gamma \cterm v\ w : \tau'} \\
  \inferrule{\Gamma \vterm v:\nat \\ \Gamma \cterm t :\tau \\ \Gamma,x:\nat \cterm t':\tau}{\Gamma \cterm \casenat{v}{t}{x}{t'} : \tau} \\
  \inferrule{\Gamma \vterm v :\tau}{\Gamma \cterm \ret{v} : \tau} \quad
  \inferrule{\Gamma \cterm t :\tau \\ \Gamma,x:\tau \cterm t :\tau'}{\Gamma \cterm \letin{x}{t}{t'} : \tau'}
\end{gathered}\]\end{minipage}}
\\[10pt]\framebox{\begin{minipage}{\linewidth}
      \emph{Operational semantics:}
      \[
  \begin{gathered}
  \inferrule{ }{\ret{v} \Downarrow v} \quad
  \inferrule{ }{\pi_1 (v,v') \Downarrow v} \quad
  \inferrule{ }{\pi_2 (v,v') \Downarrow v'} \\
  \inferrule{t[v/x] \Downarrow w}{\casesum{\inl{v}}{x}{t}{y}{t'} \Downarrow w} \\
  \inferrule{t'[v/x] \Downarrow w}{\casesum{\inr{v}}{x}{t}{y}{t'} \Downarrow w}
  \end{gathered}\qquad\begin{gathered}
  \inferrule{t[(\rec{f}{x}{t})/f,\,v/x] \Downarrow w}{(\rec{f}{x}{t})\ v \Downarrow w} \quad
  \inferrule{t[v/x] \Downarrow w}{(\lbd{x}{t})\ v \Downarrow w} \quad
  \inferrule{t \Downarrow v \\ t'[v/x] \Downarrow w}{\letin{x}{t}{t'} \Downarrow w} \\
  \inferrule{t \Downarrow w}{\casenat{\vzero}{t}{x}{t'} \Downarrow w} \quad
  \inferrule{t'[v/x] \Downarrow w}{\casenat{\suc{v}}{t}{x}{t'} \Downarrow w}\\
  \phantom{\inferrule{t \Downarrow w}{\casenat{\vzero}{t}{x}{t'} \Downarrow w} \quad
  \inferrule{t'[v/x] \Downarrow w}{\casenat{\suc{v}}{t}{x}{t'} \Downarrow w}}\end{gathered}\]\end{minipage}}
\\[10pt]
\framebox{\begin{minipage}{\linewidth}\emph{Denotational semantics of types:}
  \begin{align*}
    \abs{\den\nat} &=\bN \text{, with the discrete order} \qquad
                     \abs{\den{\tau\rightarrow \tau'}} = \omega\conc(\den\tau,\,  L_\MCM\den{\tau'}), \text{ with the pointwise order} \\
    R^c_{\den\nat} &= \big\{f:\abs{c}\rightarrow\bN \midd \exists\, \{g_i:c_i\rightarrow c\}_{i\in I}\in J(c) \text{ s.t. each } f\circ g_i \text{ is constant} \big\}\\
    R^c_{\den{\tau\rightarrow \tau'}} &= \big\{f:\abs{c}\rightarrow \omega\conc(\den\tau,\,  L_\MCM\den{\tau'}) \midd \forall h:d\rightarrow c\in\bC,\ \forall g:\abs{d}\rightarrow\abs{\den\tau} \in R^d_{\den\tau}.\ \lambda x\in\abs{d}.\big(f(h(x))\ g(x)\big)\in R^d_{L_\MCM \den{\tau'}}\big\} \\
    \abs{\den{\tau\times\tau'}} &=\abs{\den\tau}\times\abs{\den{\tau'}},\text{ where } (x,y)\leq (x',y') \text{ iff } x\leq_{\den\tau} x' \text{ and } y\leq_{\den{\tau'}}y' \\
    R^c_{\den{\tau\times\tau'}} &= \big\{\langle f,g\rangle : \abs{c}\rightarrow\abs{\den\tau}\times\abs{\den{\tau'}} \midd f \in R^c_{\den\tau},\  g\in R^c_{\den{\tau'}}\big\} \\
    \abs{\den{\tau +\tau'}} &=\abs{\den\tau} + \abs{\den{\tau'}}, \text{ where } \inl{(x)} \leq \inl{(x')} \text{ iff } x\leq_{\den\tau} x' \text{ and similarly for } \inr{} \\
    R^c_{\den{\tau+\tau'}} &=\big\{f:\abs{c}\rightarrow \abs{\den\tau} + \abs{\den{\tau'}} \midd \exists \{g_i:c_i\rightarrow c\}_{i\in I}\in J(c) \text{ s.t. for each } i,\ (f\circ g_i) \in R^{c_i}_{\den\tau} \text{ or } (f\circ g_i) \in R^{c_i}_{\den{\tau'}}\big\}
  \end{align*}\end{minipage}}
  \caption{Typing rules, operational semantics, and denotational semantics, for \pcfv.}
  \label{fig:types-interp-full}
\end{figure*}

In this appendix we provide a type system, an operational semantics, and a denotational semantics for the \pcfv\ language extended with product and sum types, as referred to in \S\ref{sec:lang}. The grammars of types, values and computations are:
\allowdisplaybreaks\begin{align*}
\tau &\Coloneqq \tzero \mid \tone \mid \nat \mid \tau + \tau \mid \tau \times \tau \mid \tau\rightarrow\tau \\
v,w &\Coloneqq x \mid \star \mid \inl{v} \mid \inr{v} \mid (v,v) \mid \vzero \mid 
  \suc{v} \mid \lbd{x}{t} \mid \rec{f}{x}{t}\\
  t  &\Coloneqq \ret{v} \mid 
                          \casesum{v}{x}{t}{y}{t'} \mid 
  \pi_1v \mid \pi_2v \\ & \mid v\ w \mid
  \casenat{v}{t}{x}{t'} \mid \letin{x}{t}{t'}   
\end{align*}
\Cref{fig:types-interp-full} provides the typing rules, operational semantics and spells out the interpretation of types for this extended language.
The big-step operational semantics of \pcfv\ is the relation $\Downarrow$ between closed computations and closed values that is the least closed under the rules. 
The interpretation of \pcfv\ types in $\omega\conc(\bC,J)$ uses the structure of the category:
\begin{gather*}
  \den\nat = 1 + 1 + \ldots \quad \den 0 = 0 \quad \den 1 = 1 \quad
  \den{\tau +\tau'}=\den\tau + \den{\tau'} \\ \den{\tau\times\tau'}=\den\tau \times \den{\tau'}\quad  \den{\tau\rightarrow \tau'}=\den{\tau}\Rightarrow L_\MCM\den{\tau'} 
\end{gather*}
The interpretation of terms uses the categorical structure in a standard way (e.g.~\cite{moggi-metalanguage}).


%% file: main.bbl
\begin{thebibliography}{10}

\bibitem{DBLP:journals/iandc/AbramskyJM00}
S.~Abramsky, R.~Jagadeesan, and P.~Malacaria.
\newblock Full abstraction for {PCF}.
\newblock {\em Inf. Comput.}, 163(2):409--470, 2000.

\bibitem{DBLP:journals/pacmpl/AguirreBGGKS21}
A.~Aguirre, G.~Barthe, M.~Gaboardi, D.~Garg, S.~Katsumata, and T.~Sato.
\newblock Higher-order probabilistic adversarial computations: categorical
  semantics and program logics.
\newblock {\em Proc. {ACM} Program. Lang.}, 5({ICFP}):1--30, 2021.

\bibitem{DBLP:conf/lics/AmorimKMPR21}
P.~H. Azevedo~de Amorim, D.~Kozen, R.~Mardare, P.~Panangaden, and M.~Roberts.
\newblock Universal semantics for the stochastic lambda calculus.
\newblock In {\em Proc.~LICS 2021}, 2021.

\bibitem{baez-hoffnung-smooth}
J.~Baez and A.~Hoffnung.
\newblock Convenient categories of smooth spaces.
\newblock {\em Trans.~AMS}, 363(11), 2011.

\bibitem{DBLP:conf/esop/BartheCLG20}
G.~Barthe, R.~Crubill{\'{e}}, U.~D. Lago, and F.~Gavazzo.
\newblock On the versatility of open logical relations - continuity, automatic
  differentiation, and a containment theorem.
\newblock In {\em Proc.~ESOP 2020}, pages 56--83, 2020.

\bibitem{DBLP:journals/pacmpl/BrunelMP20}
A.~Brunel, D.~Mazza, and M.~Pagani.
\newblock Backpropagation in the simply typed lambda-calculus with linear
  negation.
\newblock {\em Proc. {ACM} Program. Lang.}, 4({POPL}):64:1--64:27, 2020.

\bibitem{DBLP:conf/lics/Crubille18}
R.~Crubill{\'{e}}.
\newblock Probabilistic stable functions on discrete cones are power series.
\newblock In {\em Proc.~LICS 2018}, 2018.

\bibitem{DBLP:journals/pacmpl/DahlqvistK20}
F.~Dahlqvist and D.~Kozen.
\newblock Semantics of higher-order probabilistic programs with conditioning.
\newblock {\em Proc. {ACM} Program. Lang.}, 4({POPL}):57:1--57:29, 2020.

\bibitem{dubuc-concrete-quasitopoi}
E.~J. Dubuc.
\newblock Concrete quasitopoi.
\newblock In {\em Applications of Sheaves: Proceedings of the Research
  Symposium on Applications of Sheaf Theory to Logic, Algebra, and Analysis,
  Durham, July 9--21, 1977}. 1979.

\bibitem{ehrhard-concrete}
T.~Ehrhard.
\newblock On finiteness spaces and extensional presheaves over the {L}awvere
  theory of polynomials.
\newblock {\em J.~Pure Appl.~Algebra}, 2007.

\bibitem{DBLP:journals/pacmpl/EhrhardPT18}
T.~Ehrhard, M.~Pagani, and C.~Tasson.
\newblock Measurable cones and stable, measurable functions: a model for
  probabilistic higher-order programming.
\newblock {\em Proc. {ACM} Program. Lang.}, 2({POPL}):59:1--59:28, 2018.

\bibitem{escardo-xu}
M.~Escard\'o and C.~Xu.
\newblock A constructive manifestation of the {K}leene–{K}reisel continuous
  functionals.
\newblock {\em Ann.~Pure~Appl.~Logic}, 167(9):770--793, 2016.

\bibitem{fiore_1996}
M.~P. Fiore.
\newblock {\em Axiomatic Domain Theory in Categories of Partial Maps}.
\newblock Distinguished Dissertations in Computer Science. Cambridge University
  Press, 1996.

\bibitem{fiore-unpub}
M.~P. Fiore.
\newblock Enrichment and representation theorems for categories of domains and
  continuous functions.
\newblock Unpublished, 1996.

\bibitem{fiore-plotkin-adequacy}
M.~P. Fiore and G.~D. Plotkin.
\newblock An axiomatization of computationally adequate domain theoretic models
  of {F}{P}{C}.
\newblock In {\em Proc.~LICS 1994}, pages 92--102, 1994.

\bibitem{fiore-plotkin-an-extension-of-models-of-adt-to-models-of-sdt}
M.~P. Fiore and G.~D. Plotkin.
\newblock An extension of models of axiomatic domain theory to models of
  synthetic domain theory.
\newblock In {\em Computer Science Logic 1997}, pages 129--149, 1997.

\bibitem{fiore-rosolini-2sdt}
M.~P. Fiore and G.~Rosolini.
\newblock Two models of synthetic domain theory.
\newblock {\em J.~Pure~Appl.~Algebra}, 116:151--162, 1997.

\bibitem{fiore-rosolini-h}
M.~P. Fiore and G.~Rosolini.
\newblock Domains in {H}.
\newblock {\em Theoret.~Comput.~Sci.}, 264:171--193, 2001.

\bibitem{DBLP:journals/corr/abs-2106-16190}
J.~Goubault{-}Larrecq, X.~Jia, and C.~Th{\'{e}}ron.
\newblock A domain-theoretic approach to statistical programming languages.
\newblock 2021.

\bibitem{qbs}
C.~Heunen, O.~Kammar, S.~Staton, and H.~Yang.
\newblock A convenient category for higher-order probability theory.
\newblock In {\em Proc.~LICS 2017}, 2017.

\bibitem{huang_morrisett_spitters_2020}
D.~Huang, G.~Morrisett, and B.~Spitters.
\newblock An application of computable distributions to the semantics of
  probabilistic programs.
\newblock In {\em Foundations of Probabilistic Programming}, page 75–120.
  Cambridge University Press, 2020.

\bibitem{huot-staton-vakar}
M.~Huot, S.~Staton, and M.~V\'ak\'ar.
\newblock Correctness of automatic differentiation via diffeologies and
  categorical gluing.
\newblock In {\em Proc.~FOSSACS 2020}, 2020.

\bibitem{DBLP:journals/iandc/HylandO00}
J.~M.~E. Hyland and C.~L. Ong.
\newblock On full abstraction for {PCF:} i, ii, and {III}.
\newblock {\em Inf. Comput.}, 163(2):285--408, 2000.

\bibitem{diffeology-book}
P.~Iglesias-Zemmour.
\newblock {\em Diffeology}.
\newblock AMS, 2013.

\bibitem{topological-topos}
P.~Johnstone.
\newblock On a topological topos.
\newblock {\em Proc.~London Math.~Soc.}, 3(38):237--271, 1979.

\bibitem{John02}
P.~T. Johnstone.
\newblock {\em {Sketches of an elephant: a Topos theory compendium}}.
\newblock Oxford logic guides. Oxford Univ. Press, 2002.

\bibitem{DBLP:conf/nips/0001YRY20}
W.~Lee, H.~Yu, X.~Rival, and H.~Yang.
\newblock On correctness of automatic differentiation for non-differentiable
  functions.
\newblock In {\em Annual Conference on Neural Information Processing Systems
  2020, NeurIPS 2020}, 2020.

\bibitem{levy-power-thielecke}
P.~B. Levy, J.~Power, and H.~Thielecke.
\newblock Modelling environments in call-by-value programming languages.
\newblock {\em Inform.~Comput.}, 185(2):182--210, 2003.

\bibitem{DBLP:journals/pacmpl/LewCSCM20}
A.~K. Lew, M.~F. Cusumano{-}Towner, B.~Sherman, M.~Carbin, and V.~K.
  Mansinghka.
\newblock Trace types and denotational semantics for sound programmable
  inference in probabilistic languages.
\newblock {\em Proc. {ACM} Program. Lang.}, 4({POPL}):19:1--19:32, 2020.

\bibitem{DBLP:journals/corr/abs-2111-15456}
A.~K. Lew, M.~Huot, and V.~K. Mansinghka.
\newblock Towards denotational semantics of {AD} for higher-order, recursive,
  probabilistic languages.
\newblock 2021.
\newblock Presented at NeurIPS 2021 workshop on differentiable programming and
  POPL 2022 workshop on languages for inference.

\bibitem{lmz-quantum}
B.~Lindenhovius, M.~Mislove, and V.~Zamdzhiev.
\newblock Mixed linear and non-linear recursive types.
\newblock In {\em Proc.~ICFP 2019}, 2019.

\bibitem{longley-simpson-sdt-real}
J.~R. Longley and A.~K. Simpson.
\newblock A uniform approach to domain theory in realizability models.
\newblock {\em Math.~Struct.~Comput.~Sci.}, 7(5), 1997.

\bibitem{mss-presheaf-quantum}
O.~Malherbe, P.~Scott, and P.~Selinger.
\newblock Presheaf models of quantum computation: An outline.
\newblock In {\em Computation, Logic, Games, and Quantum Foundations}.
  Springer, 2013.

\bibitem{matache-moss-staton-fscd-2021}
C.~Matache, S.~Moss, and S.~Staton.
\newblock {Recursion and Sequentiality in Categories of Sheaves}.
\newblock In {\em Proc.~FSCD 2021}, volume 195, pages 25:1--25:22, 2021.

\bibitem{moggi-metalanguage}
E.~Moggi.
\newblock Notions of computation and monads.
\newblock {\em Inf.~Comput.}, 93:55--92, 1991.

\bibitem{mulry-partial-map-classifiers-and-partial-cccs}
P.~S. Mulry.
\newblock Partial map classifiers and partial cartesian closed categories.
\newblock {\em Theoretical Computer Science}, 136(1):109--123, 1994.

\bibitem{nielsen-chuang}
M.~Nielsen and I.~Chuang.
\newblock {\em Quantum Computation and Quantum Information}.
\newblock CUP, 2000.

\bibitem{DBLP:journals/iandc/OHearnR95}
P.~W. O'Hearn and J.~G. Riecke.
\newblock Kripke logical relations and {PCF}.
\newblock {\em Inf. Comput.}, 120(1):107--116, 1995.

\bibitem{plotkin-lambda-definability-and-logical-relations}
G.~Plotkin.
\newblock Lambda definability and logical relations.
\newblock Technical Report Memo SAI-RM-4, School of Artificial Intelligence,
  Edinburgh, 1973.

\bibitem{DBLP:conf/tlca/Power03}
J.~Power.
\newblock A universal embedding for the higher order structure of computational
  effects.
\newblock In {\em Proc.~TLCA 2003}, volume 2701, pages 301--315. Springer,
  2003.

\bibitem{DBLP:journals/iandc/RieckeS02a}
J.~G. Riecke and A.~Sandholm.
\newblock A relational account of call-by-value sequentiality.
\newblock {\em Inf. Comput.}, 179(2):296--331, 2002.

\bibitem{rosolini-phd}
G.~Rosolini.
\newblock {\em Continuity and effectiveness in topoi}.
\newblock PhD thesis, University of Oxford, 1986.

\bibitem{rosolini-streicher-concrete}
G.~Rosolini and T.~Streicher.
\newblock Comparing models of higher type computation.
\newblock In {\em Workshop on Realizability Semantics and Applications}, 1999.

\bibitem{DBLP:journals/pacmpl/SatoABGGH19}
T.~Sato, A.~Aguirre, G.~Barthe, M.~Gaboardi, D.~Garg, and J.~Hsu.
\newblock Formal verification of higher-order probabilistic programs: reasoning
  about approximation, convergence, bayesian inference, and optimization.
\newblock {\em Proc. {ACM} Program. Lang.}, 3({POPL}):38:1--38:30, 2019.

\bibitem{saville-kammar-katsumata-fullabs}
P.~Saville, O.~Kammar, and S.~ya~Katsumata.
\newblock Fully abstract models for effectful lambda-calculi via
  category-theoretic logical relations.
\newblock In {\em Proc.~POPL 2022}, 2022.

\bibitem{DBLP:journals/pacmpl/ScibiorKVSYCOMH18}
A.~{\'{S}}cibior, O.~Kammar, M.~V{\'{a}}k{\'{a}}r, S.~Staton, H.~Yang, Y.~Cai,
  K.~Ostermann, S.~K. Moss, C.~Heunen, and Z.~Ghahramani.
\newblock Denotational validation of higher-order bayesian inference.
\newblock {\em Proc. {ACM} Program. Lang.}, 2({POPL}):60:1--60:29, 2018.

\bibitem{simpson-computational-adequacy-in-an-elementary-topos}
A.~K. Simpson.
\newblock Computational adequacy in an elementary topos.
\newblock In G.~Gottlob, E.~Grandjean, and K.~Seyr, editors, {\em Computer
  Science Logic}, pages 323--342, Berlin, Heidelberg, 1999. Springer Berlin
  Heidelberg.

\bibitem{DBLP:journals/apal/Simpson04}
A.~K. Simpson.
\newblock Computational adequacy for recursive types in models of
  intuitionistic set theory.
\newblock {\em Ann. Pure Appl. Log.}, 130(1-3):207--275, 2004.

\bibitem{smootheology}
A.~Stacey.
\newblock Comparative smootheology.
\newblock {\em Theory Appl.~Categ.}, 25(4):64--117, 2011.

\bibitem{sterling-harper-sheaf-semantics-of-termination-insensitive-noninterference}
J.~Sterling and R.~Harper.
\newblock Sheaf semantics of termination-insensitive noninterference.
\newblock In A.~Felty, editor, {\em 7th International Conference on Formal
  Structures for Computation and Deduction (FSCD 2022)}, volume 228 of {\em
  Leibniz International Proceedings in Informatics (LIPIcs)}, Dagstuhl,
  Germany, Aug. 2022. Schloss Dagstuhl--Leibniz-Zentrum fuer Informatik.

\bibitem{DBLP:journals/corr/abs-2007-05282}
M.~V{\'{a}}k{\'{a}}r.
\newblock Denotational correctness of foward-mode automatic differentiation for
  iteration and recursion.
\newblock {\em CoRR}, abs/2007.05282, 2020.

\bibitem{DBLP:conf/esop/Vakar21}
M.~V{\'{a}}k{\'{a}}r.
\newblock Reverse {AD} at higher types: Pure, principled and denotationally
  correct.
\newblock In {\em Proc.~ESOP 2020}, pages 607--634, 2021.

\bibitem{DBLP:journals/pacmpl/VakarKS19}
M.~V{\'{a}}k{\'{a}}r, O.~Kammar, and S.~Staton.
\newblock A domain theory for statistical probabilistic programming.
\newblock {\em Proc. {ACM} Program. Lang.}, 3({POPL}):36:1--36:29, 2019.

\bibitem{winskel-semantics}
G.~Winskel.
\newblock {\em The Formal Semantics of Programming Languages: An Introduction}.
\newblock MIT Press, 1993.

\bibitem{DBLP:conf/aistats/ZhouGKRYW19}
Y.~Zhou, B.~J. Gram{-}Hansen, T.~Kohn, T.~Rainforth, H.~Yang, and F.~Wood.
\newblock {LF-PPL:} {A} low-level first order probabilistic programming
  language for non-differentiable models.
\newblock In {\em Proc.~AISTATS 2019}, pages 148--157, 2019.

\end{thebibliography}
